\documentclass[reqno,12pt,a4paper]{article}
\usepackage{amsmath}
\usepackage{amssymb}
\usepackage{amsfonts}
\usepackage{float}
\usepackage[font=small,labelfont=bf,margin=10pt,
    justification=centering, format=default,
    singlelinecheck=false]{caption}
\usepackage[
    plainpages=false,
    bookmarks=true,
    bookmarksopen,
    colorlinks=true,
    urlcolor=blue]
{hyperref}
\usepackage{latexsym}
\usepackage{titlefoot}
\usepackage[sort,longnamesfirst]{natbib}
\usepackage{geometry}
\usepackage{setspace}
\usepackage{pdflscape}
\usepackage{titlesec}
\usepackage{colortbl}
\usepackage{enumitem}

\makeatletter
    \newcommand\floatc@simple[2]{{\@fs@cfont #1} #2\par}
    \newcommand\fs@onerule{\def\@fs@cfont{\bfseries}\let\@fs@capt\floatc@simple
        \def\@fs@mid{\hrule \kern8pt}%
        \def\@fs@pre{}%
        \def\@fs@post{}%
        \let\@fs@iftopcapt\iftrue}
\makeatother

\newcommand{\qed}{\quad\hbox{\vrule width 4pt height 5pt depth 1.5pt}}
\newcommand{\half}{\frac{1}{2}}

\makeatletter
\newlength{\myFootnoteWidth}
\newlength{\myFootnoteLabel}
\renewcommand{\@makefntext}[1]{%
  \setlength{\myFootnoteWidth}{\columnwidth}%
  \addtolength{\myFootnoteWidth}{-\myFootnoteLabel}%
  \noindent\makebox[\myFootnoteLabel][r]{\@makefnmark\ }%
  \parbox[t]{\myFootnoteWidth}{#1}%
}
\makeatother

\makeatletter
\renewcommand\footnoterule{%
  \vspace{1em}
  \kern-3\p@\hrule\@width.4\columnwidth%
  \kern3.6\p@}
\makeatother

\titleformat{\section}{\normalfont\Large\bfseries\singlespacing}{\thesection}{1em}{}
\titlespacing*{\section}{0pt}{0.2\baselineskip}{\baselineskip}
\floatstyle{onerule}
\restylefloat{table}
\newtheorem{theorem}{Theorem}
\newtheorem{assumption}{Assumption}
\newtheorem{lemma}{Lemma}
\newtheorem{proposition}{Proposition}
\definecolor{hcolor}{rgb}{0.85,0.85,0.85}
\mathchardef\mhyphen="2D
\setlist[enumerate,1]{label=\bfseries\arabic*., leftmargin=2em}
\setlist[enumerate,2]{label=\bfseries(\alph*), leftmargin=1.5em}
\let \oldmarginpar \marginpar
\renewcommand\marginpar[1]{\-\oldmarginpar[\raggedleft\footnotesize #1]{\raggedright\color{blue}\singlespacing\tiny\vspace{-1cm} #1}}
\hypersetup{pdfstartview={FitH}, citecolor=blue, linkcolor=blue}

\begin{document}

\title{Bias Correction of Semiparametric Long Memory Parameter Estimators
via the Pre-filtered Sieve Bootstrap\thanks{%
This research has been supported by Australian Research Council (ARC)
Discovery Grant DP120102344. The authors would like to thank the Editor, a
co-editor and two referees for very detailed and constructive comments on
earlier drafts of the paper.}}
\author{D. S. Poskitt, Gael M. Martin\thanks{%
Corresponding author: Gael Martin, Department of Econometrics and Business
Statistics, Monash University, Victoria 3800, Australia. \emph{Tel.:}%
+61-3-9905-1189; \emph{fax:}+61-3-9905-5474; \emph{E-mail:}
gael.martin@monash.edu. } and Simone D. Grose \\
{\small \emph{Department of Econometrics \& Business Statistics, Monash
University}}}
\maketitle

\begin{abstract}
This paper investigates bootstrap-based bias correction of semiparametric
estimators of the long memory parameter, $d$, in fractionally integrated
processes. The re-sampling method involves the application of the sieve
bootstrap to data pre-filtered by a preliminary semiparametric estimate of
the long memory parameter. Theoretical justification for using the bootstrap
technique to bias adjust log periodogram and semiparametric local Whittle
estimators of the memory parameter is provided in the case where the true
value of $d$ lies in the range $0\leq d<0.5$. That the bootstrap method
provides confidence intervals with the correct asymptotic coverage is also
proven, with the intervals shown to adjust explicitly for bias, as estimated
via the bootstrap. Simulation evidence comparing the performance of the
bootstrap bias correction with analytical bias-correction techniques is
presented. The bootstrap method is shown to produce notable bias reductions,
in particular when applied to an estimator for which some degree of bias
reduction has already been accomplished by analytical means.
\end{abstract}

\setcounter{footnote}{1}

\noindent {\footnotesize \emph{MSC2010 subject classifications}: Primary
62M10, 62M15; Secondary 62G09}


\thispagestyle{empty} \setcounter{page}{0}

\section{Introduction}

The so-called long memory, or strongly dependent, processes have come to
play an important role in time series analysis. Long-range dependence,
observed in a very wide range of empirical applications, is characterized by
an autocovariance structure that decays too slowly to be absolutely
summable. Specifically, rather than the autocovariance function declining at
the exponential rate characteristic of a stable and invertible ARMA process,
it declines at a hyperbolic rate dependent on a \textquotedblleft long
memory\textquotedblright\ parameter. A detailed description of the
properties of such processes can be found in \cite{beran:1994}. Perhaps the
most popular model of a long memory process is the fractionally integrated ($%
I(d)$) process introduced by \cite{granger:joyeux:1980} and \cite%
{hosking:1981}. This class of processes can be characterized by the
specification,
\begin{equation}
y(t)=\sum_{j=0}^{\infty }k(j)\varepsilon (t-j)=\frac{\kappa (z)}{(1-z)^{d}}%
\,\varepsilon (t),  \label{Wold}
\end{equation}%
where $\varepsilon (t)$ is zero-mean white noise, $z$ is here interpreted as
the lag operator $(z^{j}y(t)=y(t-j))$, and $\kappa (z)=\sum_{j\geq 0}\kappa
(j)z^{j}$, $\kappa (0)=1$. For any $d>-1$ the operator $(1-z)^{d}$ is
defined via a binomial expansion and if the \textquotedblleft short
memory\textquotedblright\ component $\kappa (z)$ is the transfer function of
a stable, invertible ARMA process and $|d|<0.5$, then the coefficients of $%
k(z)$ are square-summable $(\sum_{j\geq 0}|k(j)|^{2}<\infty )$. In this case
$y(t)$ is well-defined as the limit in mean square of a
covariance-stationary process and the model is essentially a generalization
of the classic Box-Jenkins ARIMA model \citep{box:jenkins},
\begin{equation}
(1-z)^{d}\Phi (z)y(t)=\Theta (z)\varepsilon (t),  \label{eq:arfima}
\end{equation}%
in which we now allow non-integer values of the integrating parameter $d$
and $\kappa (z)=\Theta (z)/\Phi (z).$

The long-run behaviour of the process in (\ref{eq:arfima}) naturally depends
on the fractional integration parameter $d$. In particular, for any $d>0$
the impulse response coefficients of the Wold representation in \eqref{Wold}
are not absolutely summable and, for $0<d<0.5$, the autocovariances decline
at the rate $\gamma (\tau )\sim C\tau ^{2d-1}$. Such processes have been
found to exhibit dynamic behaviour very similar to that observed in many
empirical time series. See \cite{robinson:2003} for a collection of the
seminal articles in the area and \cite{doukhan:oppenheim:taqqu:2003} for a
thorough review of theory and applications.

Statistical procedures for analyzing long memory processes have ranged from
the likelihood-based methods of \cite{fox:taqqu:1986}, \cite{dahlhaus:1989},
\cite{sowell:1992} and \cite{beran:1995}, to the semiparametric techniques
advanced by \cite{geweke:porter:1983} and \cite%
{robinson:1995a,robinson:1995b}, among others. The asymptotic theory for
maximum likelihood estimation of the parameters of such processes is well
established, at least under the assumption of Gaussian errors. In
particular, we have consistency, asymptotic efficiency, and asymptotic
normality for the MLE of the fractional differencing parameter, so providing
a basis for large sample inference in the usual manner. Such asymptotic
results are, however, conditional on correct model specification, with the
MLE of $d$ typically inconsistent if either or both the autoregressive and
moving average operators in \eqref{eq:arfima} (or, equivalently, the
operator $\kappa (z)$ in \eqref{Wold}) are incorrectly specified. The
semiparametric methods aim to produce consistent estimators of $d$ while
placing only very mild restrictions on the behaviour of $\kappa (e^{\imath
\lambda })$ for frequency values $\lambda $ near zero. The semiparametric
estimators are therefore robust to different forms of short-run dynamics and
offer broader applicability than a fully parametric method. They are also
asymptotically pivotal and have particularly simple asymptotic normal
distributions.

Whilst such features place the semiparametric methods at the forefront for
use in conducting inference on $d$, the price paid for their application is
a reduction in asymptotic efficiency (relative to exact ML) and a slower
rate of convergence to the true parameter %
\citep{giraitis:robinson:samarov:1997}. Also, despite asymptotic robustness
to the short-run dynamics, semiparametric estimators have been shown to
exhibit large finite sample bias in the presence (in particular) of a
substantial autoregressive component -- see \cite%
{agiakloglou:newbold:wohar:1993} and \cite{lieberman:2001b} for examples. In
response to these findings, analytical approaches to reducing the
first-order bias of semiparametric estimators have been proposed. \cite%
{moulines:soulier:1999}, for example, reduce bias by fitting a finite number
of Fourier coefficients to the logarithm of the short memory spectrum and
constructing a broad-band log periodogram regression (LPR) estimator of $d$\
that uses all of the frequencies in the range $(0,\pi ]$, not just those in
a neighborhood of zero. \cite{andrews:guggenberger:2003} consider a
bias-adjusted estimator of $d$\ obtained by including even powers of
frequency as additional regressors in the pseudo regression that defines the
LPR estimator, and \cite{andrew:sun:2004} adapt this approach to the
semiparametric local Whittle (SPLW) estimator examined in \cite%
{robinson:1995b}. 

As a point of contrast with existing work, the focus of this paper is on the
use of the bootstrap to bias correct semiparametric estimators of the long
memory parameter. As is consistent with the semiparametric approach to
estimation of $d$ itself, a semiparametric approach to the bootstrap scheme
is also adopted, based on the \textquotedblleft sieve\textquotedblright\
technique. This works by \textquotedblleft pre-whitening\textquotedblright\
the data using an autoregressive approximation, with the dynamics of the
process captured in a fitted autoregression \citep[see][]{politis:2003}.
Provided the order, $h$, of the autoregression increases at a suitable rate
with $T$, the convergence rates for the sieve bootstrap are much closer (in
fact arbitrarily close) to those for simple random samples. \cite%
{choi:hall:2000} demonstrate the superior convergence performance of the
sieve bootstrap over the block bootstrap for linear short memory processes,
while \citet{poskitt:grose:martin:2015} build on the results of \cite%
{poskitt:2008} to show that under regularity conditions that allow for
$I(d)$\ processes the sieve bootstrap achieves an error rate of $%
O_{p}(T^{-(1-\max \{0,d\})+\beta })$, $\beta >0$, for the quantiles of the
sampling distribution of a general class of statistics that includes the
sample mean and second-order moments.

The current paper uses a modified sieve bootstrap, wherein a consistent
semiparametric estimator of the long memory parameter is used to pre-filter
the raw data, prior to the use of a long autoregressive approximation as the
sieve from which bootstrap samples are produced. As the focus of the paper
is on the use of the bootstrap to bias correct, theoretical results are
presented that pertain directly to the accuracy with which the pre-filtered
sieve method estimates the true bias in the relevant estimators of $d.$
Specifically, we derive error rates for bootstrap-based estimation of the
bias of $\sqrt{N}$--CAN (consistent and asymptotically normal); $N\sim
KT^{\nu }$, $K\in (0,\infty )$, $0<\nu <1$; estimators that satisfy a
requisite Edgeworth expansion, subject to the pre-filtering value itself
converging almost surely to the true value of $d$ at a sufficient rate. The
theoretical validity of bootstrap highest probability density (HPD)
confidence intervals constructed from the pre-filtered bootstrap
replications is also established. To demonstrate the bootstrap technique we
use it to bias correct the LPR and SPLW estimators, plus the
analytically-bias-adjusted variants of \cite{andrews:guggenberger:2003} and
\cite{andrew:sun:2004}.

Our exposition centers around the short and long memory stationary case,
with the true value of $d\, $ assumed to lie in the range $0\leq d<0.5$.
Whilst this may be deemed to be a limitation of sorts, our key theoretical
results are stated in a form that suggests that they will have more general
applicability, subject only to the proviso that the assumption of
stationarity can be relaxed to accommodate more general processes. For
example, non-stationary long memory structures could be catered for by
considering data generating mechanisms driven by fractional noise of the
form
\begin{equation*}
n(t)=\left\{
\begin{array}{ll}
\sum_{s=0}^{t-1}\alpha _{s}^{(d)}\varepsilon (t-s), & d\in \lbrack 0,0.5)\,;
\\
\sum_{s=0}^{t-1}\sum_{\tau =0}^{t-s-1}\alpha _{\tau }^{(1-d)}\varepsilon
(t-s-\tau ), & d\in \lbrack 0.5,1.5)\,,%
\end{array}%
\right.
\end{equation*}%
where $\alpha _{s}^{(d)}$, $s=0,1,2,\ldots $, denote the coefficients of the
fractional difference operator when expressed in terms of its binomial
expansion, as in equation \eqref{pocham} below. The pre-filtered sieve
bootstrap could then be applied as described in Section 2 using an
appropriate estimator, such as the quasi (Gaussian) maximum likelihood
estimator of \citet{tanaka:1999} or the exact local Whittle estimator of %
\citet{shimotsu:phillips:2005}. The difficulty here lies not in the
practical implementation of the pre-filtered sieve bootstrap for such
estimators, but in showing that a large-deviations condition necessary to
establish the theoretical validity of the method holds -- we will return to
a brief discussion of this issue below.

In addition to the theoretical results, an extensive simulation exercise is
undertaken, in which the bias and mean squared error (MSE) of the
bootstrap-bias-adjusted estimators is documented, in comparison with the
corresponding statistics both for the original unadjusted estimators, and
the estimators that are adjusted by analytical methods alone. As a benchmark
for the effectiveness of the bias correction we also present bias and MSE
results for the correctly specified (and hence asymptotically optimal) MLE.
The bootstrap bias adjustment is implemented both as a one-step exercise,
and as an iterative procedure, with a stochastic stopping rule invoked to
produce the final estimator. The empirical coverage (and average length) of
the HPD confidence intervals is also recorded for all estimators. In
accordance with the theoretical results, we investigate the (relative)
performance of the bootstrap bias adjustment using values of $d$ in the
range $0\leq d<0.5$ to generate the data in the simulation experiments. The
LPR and SPLW estimators themselves however -- both in unadjusted and
bias-adjusted form - are essentially left unconstrained, as there is nothing
in the structure of the pre-filtered sieve bootstrap algorithm per se that
requires that the estimator that is to be bias corrected, or the pre-filter,
be restricted to lie in the $0$ to $0.5$ range.\footnote{%
The qualification `essentially' contained in this statement refers to a
deterministic stopping criterion that supplements two stochastic stopping
rules applied, in turn, to the iterative version of the bias-correction
method. This point is discussed further in Sections 4 and 5.1.}

The paper proceeds as follows. Section 2 briefly summarizes the statistical
properties of long memory processes, and outlines the pre-filtered sieve
bootstrap in this context. The bootstrap-based bias-adjustment algorithm is
also described in this section. In Section 3 we present the key theoretical
results, namely the almost sure convergence of the bootstrap estimator of
the true bias, and the correctness of the (asymptotic) coverage probability
of the bootstrap confidence intervals. The associated proofs are assembled
in Appendix \ref{proofs}. Section 4 outlines the iterated version of the
bootstrap bias-adjustment technique, with details of the stochastic stopping
rules invoked therein given in Appendix \ref{ssr}. Details and discussion of
the simulation study follow in Section 5, the results of which are tabulated
in Appendix \ref{tbls}. Section 6 concludes the paper with a
summary of our contribution and a discussion of the extension of the
pre-filtered sieve bootstrap methodology to more general processes.

\bigskip

\section{Long memory Processes, Autoregressive Approximation, and the
Pre-Filtered Sieve Bootstrap}

Let $y(t)$ for $t\in {\mathcal{Z}}$ denote a linearly regular,
covariance-stationary process, with representation as in \eqref{Wold}, where:

\begin{assumption}
\label{Ass1} The transfer function in the representation \eqref{Wold} is
given by $k(z)=\kappa(z)/(1-z)^d$ where $d\in[0,0.5)$ and $\kappa(z)\neq 0$,
$|z|\leq 1$. The impulse response coefficients of $\kappa(z)$ satisfy $%
k(0)=1 $ and $\sum_{j\geq 0}j|\kappa(j)|<\infty$.
\end{assumption}

\begin{assumption}
\label{Ass2} The innovations process $\varepsilon(t)$ is an $i.i.d.$ zero
mean Gaussian white noise process with variance $\sigma^2$.
\end{assumption}

Assumption \ref{Ass1} serves to characterize the spectral features of quite
a wide class of short and long memory processes, including long-range
dependent members of the ARFIMA family of models that are the focus of this
paper. This assumption implies that the innovations in \eqref{Wold} are
\emph{fundamental}; meaning that $\varepsilon (t)$ lies in the space spanned
by current and past values of $y(t)$, and $\varepsilon (t)$ and $y(s) $ are
uncorrelated for all $s<t$. For a discussion of the role of fundamentalness
in the context of the autoregressive sieve bootstrap see %
\citet{kreiss:paparoditis:politis:2011}. Note that the regularity conditions
employed in \citet{kreiss:paparoditis:politis:2011} exclude fractional time
series, but using the extension of Baxter's inequality to long-range
dependent processes due to \citet{inoue:kasahara:2006} it is possible to
generalize the results of \citet{kreiss:paparoditis:politis:2011} to time
series generated from a fractional transformation of a linear process. In
particular, since the statistics that we investigate are asymptotically
pivotal the results in \citet[][Section 3]{kreiss:paparoditis:politis:2011}
can be extended to the statistics and class of processes under consideration
here.

Assumptions \ref{Ass1} and \ref{Ass2} imply that $y(t)$ is a Gaussian linear
process. A basic property of a linear process that underlies the sieve
bootstrap methodology and the associated results is that $y(t)$ is linearly
regular and the linear predictor $\bar{y}(t)=\sum_{j=1}^{\infty }\pi
(j)y(t-j)\,,$ where $\sum_{j=0}^{\infty }\pi (j)z^{j}=(1-z)^{d}\kappa
(z)^{-1}$, is the minimum mean squared error predictor of $y(t)$ based upon
its entire past. The need to invoke Gaussianity is unfortunate but is
unavoidable here as we wish to employ certain results from the existing
literature where the assumption that $y(t)$ is a Gaussian process is adopted. The use
of these results is made explicit in Section 3.
It is likely that our results can be extended to more general linear
processes, although the regularity conditions and manipulations needed for
such extensions are liable to be relatively involved. \cite%
{fay:moulines:soulier:2004}, for example, provide a discussion of Edgeworth
expansions in the context of linear statistics applied to long-range
dependent linear processes where the innovations process is $i.i.d.$ zero
mean white noise with variance $\sigma ^{2}$, but Gaussianity is replaced by
a strengthening of the Cram\'{e}r condition on the characteristic function
of the innovations. Extensions of the results in \cite%
{fay:moulines:soulier:2004} to the LPR estimator are presented in %
\citet{fay:2010}.\footnote{%
Edgeworth expansions for quadratic forms and the MLE in Gaussian long memory
series are developed in
\citet{lieberman:rousseau:zucker:2001,
lieberman:rousseau:zucker:2003}. See also %
\citet{lieberman:rosemarin:rousseau:2011}.}

The minimum mean squared error predictor of $y(t)$ based only on a finite
number $h$ of past observations ($\mathit{MMSEP}(h)$) is $\bar{y}%
_{h}(t)=\sum_{j=1}^{h}\pi _{h}(j)y(t-j)\equiv -\sum_{j=1}^{h}\phi
_{h}(j)y(t-j),$ where the minor reparameterization from $\pi _{h}$ to $\phi
_{h}$ allows us, on also defining $\phi _{h}(0)=1$, to conveniently write
the corresponding prediction error in the form of an autoregression of order
$h$ ($\textit{AR}(h)$), namely $\varepsilon _{h}(t)=\sum_{j=0}^{h}\phi _{h}(j)y(t-j).$
The finite-order autoregressive coefficients $\phi _{h}(1),\ldots ,\phi
_{h}(h)$ can be derived from the Yule-Walker equations $\sum_{j=0}^{h}\phi
_{h}(j)\gamma (j-k)=\delta _{0}(k)\sigma _{h}^{2}\,,\quad k=0,1,\ldots ,h$;
in which $\gamma (\tau )=\gamma (-\tau )=E[y(t)y(t-\tau )]$, $\tau
=0,1,\ldots $, is the autocovariance function of the process $y(t)$, $\delta
_{0}(k)$ is Kronecker's delta (i.e., $\delta _{0}(k)=0\;\forall \;k\neq
0;\;\delta _{0}(0)=1$), and the minimum mean squared error is $\sigma
_{h}^{2}=E\big[\varepsilon _{h}(t)^{2}\big]\,$, the prediction error
variance associated with $\bar{y}_{h}(t)$.

The use of autoregressive models of finite order $h$ to approximate an
unknown (but suitably regular) process therefore requires that the optimal
predictor $\bar{y}_{h}(t)$ determined from the $\textit{AR}(h)$ model be a good
approximation to the \textquotedblleft infinite-order\textquotedblright\
predictor $\bar{y}(t)$ for sufficiently large $h$. The asymptotic validity
and properties of $\textit{AR}(h)$ models when $h\rightarrow \infty $ with the sample
size $T$, under regularity conditions that admit long-range dependent
processes, were established in \cite{poskitt:2007}, and we refer the reader
to that paper for more details. That the sieve bootstrap, which uses an $%
AR(h)$ approximation to capture the dynamics of $y(t)$ (with $h$ selected
optimally) is, accordingly, a plausible semiparametric bootstrap for a long
memory process, was subsequently established in \cite{poskitt:2008}. We
focus in this paper on a modified version of this form of bootstrap.%
\footnote{\citet{andrews:lieberman:marmer:2006} examine properties of the
parametric bootstrap for the current class of processes. Our aim in this
exercise, however, is to avoid full parametric specifications and the
associated implications of misspecification. Recent (non-bootstrap-based)
work in \citet{nadarajah:martin:poskitt:2014} indicates that substantial
bias can be incurred by various parametric estimators, including the
Gaussian MLE, as a result of misspecification, highlighting that the nature
of any misspecification would be critical to the performance of associated
parametric bootstrap procedures.}

\subsection{The pre-filtered sieve bootstrap\label{pfs}}

Let $\alpha _{j}^{(d)}$; $d\in \lbrack 0,0.5)$; $j=0,1,2,\ldots $; denote
the coefficients of the binomial expansion of the fractional difference
operator $(1-z)^{d}=\sum_{j=0}^{\infty }\alpha _{j}^{(d)}z^{j}$,%
\begin{equation}  \label{pocham}
\alpha _{j}^{(d)}=\frac{\Gamma (j-d)}{\Gamma (-d)\Gamma (j+1)};\text{ }%
j=0,1,2,\ldots
\end{equation}
The pre-filtered sieve bootstrap (PFSB) realizations of $y(t)$ are generated
using the following algorithm:

\begin{enumerate}
\item \label{step1} For a given preliminary value $d^{f}$ of $d$ calculate
the coefficients of the filter $(1-z)^{d^{f}}$, and from the observed data
generate the filtered series
\begin{equation}
w^{f}(t)=\sum_{j=0}^{t-1}\alpha _{j}^{(d^{f})}y(t-j)\,,\quad t=1,\ldots ,T\,.
\label{wft}
\end{equation}

\item \label{step2} Fit an autoregression to $w^{f}(t)$ and generate a sieve
bootstrap sample $w^{\ast _{f}}(t)$, $t=1,\ldots ,T$, of the filtered data
as follows:

\begin{enumerate}
\item Given the filtered series $w^{f}(t)$, $t=1,\ldots ,T$, calculate the
parameter estimates, $\bar{\phi}_{h}(1),\ldots ,\bar{\phi}_{h}(h)$ and $\hat{%
\sigma}_{h}^{2}$, of the $\textit{AR}(h)$ approximation, and evaluate the residuals $%
\bar{\varepsilon}_{h}(t)=\sum_{j=0}^{h}\bar{\phi}_{h}(j)w^{f}(t-j)$, $%
t=1,\ldots ,T$, using $w^{f}(1-j)=w^{f}(T-j+1)$, $j=1,\ldots ,h$, as initial
values.

\item Construct the standardized residuals $\tilde{\varepsilon}_{h}(t)=(\bar{%
\varepsilon}_{h}(t)-\bar{\varepsilon}_{h})/\bar{\sigma}_{h}$, $t=1,\ldots ,T$%
, where $\bar{\varepsilon}_{h}=T^{-1}\sum_{t=1}^{T}\bar{\varepsilon}_{h}(t)$
and $\bar{\sigma}_{h}^{2}=T^{-1}\sum_{t=1}^{T}(\bar{\varepsilon}_{h}(t)-\bar{%
\varepsilon}_{h})^{2}$.%

\item Set $\varepsilon _{h}^{\ast }(t)=\bar{\sigma}_{h}e(t)$, $t=1,\ldots ,T$%
, where $e(t)$, $t=1,\ldots,T$, denotes a simple random sample of $T$
\textit{i.i.d.} values drawn from the standard normal distribution.

\item Construct the sieve bootstrap realization $w^{\ast _{f}}(1),\ldots
,w^{\ast _{f}}(T)$ using the autoregressive process $\sum_{j=0}^{h}\bar{\phi}%
_{h}(j)w^{\ast _{f}}(t-j)=\varepsilon _{h}^{\ast }(t)$, $t=1,\ldots ,T$,
initiated at $w^{\ast _{f}}(1-j)=w^{f}(\tau -j+1)$, $j=1,\ldots ,h$, where $%
\tau $ has the discrete uniform distribution on the integers $h,\ldots ,T$.
\end{enumerate}

\item Using the coefficients of the (inverse) filter $(1-z)^{-d^{f}}$,
construct the corresponding pre-filtered sieve bootstrap draw of the process
$y^{\ast _{f}}(t)=\sum_{j=0}^{t-1}\alpha _{j}^{(-d^{f})}w^{\ast _{f}}(t-j)$,
$t=1,\ldots ,T$.
\end{enumerate}

The basic, or `raw', sieve bootstrap is equivalent to setting $d^{f}=0$ in
the {PFSB}; in which case Steps 1 and 3 are redundant and Step 2 is applied
to the raw data $y(t)$.
The properties of the raw sieve bootstrap for fractional processes are given
in \cite{poskitt:2008}. Crucially, if $\bar{\phi}_{h}(z)=\sum_{j=0}^{h}\bar{%
\phi}_{h}(j)z^{j}$ denotes the estimator of $\phi
_{h}(z)=\sum_{j=0}^{h}\phi_{h}(j)z^{j}$ when the sieve bootstrap is applied
to the raw data, and $y(t)$ is a linearly regular, covariance-stationary
process that satisfies Assumptions \ref{Ass1} and \ref{Ass2}, then for all $%
h\leq H_{T}=a(\log T)^{c}$, $a>0$, $c<\infty $, $\sum_{j=1}^{h}|\bar{\phi}%
_{h}(j)-\phi _{h}(j)|^{2}=O\left( h(\log T/T)^{1-2\,max\{0,d\}}\right) $
a.s. (See also Theorem 5 and Corollary 1 of \citealp{poskitt:2007}, and the
associated discussion.) Given that the order of magnitude of $|\bar{\phi}%
_{h}(z)-\phi _{h}(z)|$ is a function of the fractional integration
parameter, it is apparent that convergence of the algorithm must depend on
the values of $d^{f}$ and $d$.\footnote{\citet{poskitt:grose:martin:2015}
build on \cite{poskitt:2008} to show that under appropriate conditions, and
for particular statistics, a sieve bootstrap generated sampling distribution
achieves a convergence rate of $O_{p}(T^{-(1-\max \{0,d\})+\beta })$ for all
$\beta >0$ and $|d|<0.5$. Obviously, the closer is $d$ to zero the closer
the convergence rate will be to $O_{p}(T^{-1+\beta })$, the rate achieved
with short memory (and anti-persistent) processes, and a rate arbitrarily
close to that achieved with simple random samples.}

Now, when $d^{f}\neq 0$, $(1-z)^{d^{f}}y(t)=(1-z)^{d^{f}-d}\kappa
(z)\varepsilon (t)$ has fractional index $d-d^{f}$, where by assumption $%
|d^{f}-d|$ $=o(1)$ $a.s.$; i.e. the pre-filtering value -- presumed to be
estimated from the data, and denoted hereafter by $d^{f}=d_{T}^{f}$
accordingly -- is assumed to be a strongly consistent estimator of $d.$
Hence, for any $\delta >0$ the event $(d_{T}^{f}-d)\in N_{\delta
}=\{x:|x|<\delta \}$ will occur with probability one as $T\rightarrow \infty
$. The error in the AR approximation to $w^{f}(t)$ will accordingly be of
order $O(h(\log T/T)^{1-2\delta })$ or smaller. In Section 3 it is shown
that $d_{T}^{f}$ needs to satisfy the large deviations property $%
|d_{T}^{f}-d|\log T\rightarrow 0$ a.s. as $T\rightarrow \infty $ in order
for this level of accuracy to be transferred to the pre-filtered sieve
bootstrap realizations $y^{\ast _{f}}(t)$ of $y(t)$, via the sieve bootstrap
draws $w^{\ast _{f}}(t)$ of $w^{f}(t)$. Theoretical results pertaining to
the accuracy of the PFSB algorithm as a mechanism for bias reduction of
semiparametric long memory parameter estimators are then provided.

\subsection{Bias correction via the pre-filtered sieve bootstrap\label{bc}}

With the conditions on $d_{T}^{f}$ verified in any particular case, we
employ the pre-filtered sieve bootstrap for the purpose of bias correcting
the LPR and SPLW estimators of the memory parameter, and their analytically
adjusted variants.
To bias correct any chosen estimator, $\widehat{d}_{T}$, of $d$ we proceed
as follows:

\begin{enumerate}
\item Calculate $\widehat{d}_{T}$ from the data $y(t)$, $t=1,\ldots ,T$.

\item Using an appropriate data-based pre-filtering value $d^{f}=d_{T}^{f}$,
produce $B$ bootstrap realizations $y_{b}^{\ast _{f}}(t)$, $t=1,\ldots ,T$; $%
b=1,2,\ldots ,B$; of the process $y(t)$. From these construct $B$ bootstrap
values of the estimator, $\widehat{d}_{T,b}^{\ast _{f}}$, $b=1,2,...,B$, by
evaluating the estimator $\widehat{d}_{T}$ for each of the $B$ independent
bootstrap draws.

\item Estimate the bias of $\widehat{d}_{T}$ by
\begin{equation}
\widehat{b}_{T,B}^{\ast _{f}}=\overline{d}_{T,B}^{\ast _{f}}-d_{T}^{f}
\label{bs_bias}
\end{equation}%
where
\begin{equation}
\overline{d}_{T,B}^{\ast _{f}}=B^{-1}\sum\limits_{b=1}^{B}\widehat{d}%
_{T,b}^{\ast _{f}}  \label{dstar_bar}
\end{equation}%
and produce the bias-adjusted estimator
\begin{equation}
\widetilde{d}_{T}=\widehat{d}_{T}-\widehat{b}_{T,B}^{\ast _{f}}.
\label{bias_adjusted_est}
\end{equation}
\end{enumerate}

\bigskip

While there is no fundamental requirement that the pre-filtering value
correspond to the estimator being bias corrected, this correspondence is a
natural one to adopt. As such, $d_{T}^{f}$ is initially taken to represent
either the LPR or SPLW estimator, with or without analytical bias
adjustment, according to whichever of these estimators is the subject of
bias correction. With the introduction of an iterative version of this
bias-correction procedure in Section \ref{recursive}, the set of pre-filters
is expanded to include bootstrap-based bias-corrected versions of the base
estimators. The validity of all such versions of the pre-filter is
established in the following section.

\section{Key Theoretical Results\label{theory}}

\subsection{Convergence of the bootstrap estimator of bias}

Motivated by a consideration of the properties of the LPR and SPLW
estimators that are the focus herein, suppose that $\widehat{d}_{T}$ (the
estimator to be bias corrected) is an asymptotically pivotal $\sqrt{N}$--CAN
estimator of $d$ where $N\sim KT^{\nu }$, $K\in (0,\infty )$, and, following %
\citet{hurvich:deo:brodsky:1998} and \cite{giraitis:robinson:2003}, wherein
Gaussianity is assumed as it is in the current paper, we now restrict $\nu $
to lie in the range $(2/3,4/5)$. Let $b_{T}$ denote the finite sample bias
of $\widehat{d}_{T}$, that is,
\begin{equation}
b_{T}=E[\widehat{d}_{T}]-d.  \label{bt}
\end{equation}%
Here $E$ denotes expectation taken with respect to the original probability
space $(\Omega ,\mathfrak{F},P)$. Now let $\widehat{d}_{T}^{\ast _{f}}$
denote the value of $\widehat{d}_{T}$ calculated from a bootstrap
realization of the process, $y^{\ast _{f}}(t)$, $t=1,\ldots ,T$, constructed
using the PFSB algorithm where the pre-filtering value $d_{T}^{f}$ by
construction satisfies the conditions stated above for $\widehat{d}_{T}$,
given that we equate $d_{T}^{f}$ to $\widehat{d}_{T}$ in any particular
instance. Since the process $\varepsilon _{h}^{\ast }(t)$ is Gaussian, it follows that
$y^{\ast _{f}}(t)$ will be a fractionally integrated $\textit{AR}(h)$ Gaussian
process with parameters $d_{T}^{f}$ and $\bar{\phi}_{h}(1),\ldots ,\bar{\phi}%
_{h}(h)$, and $\widehat{d}_{T}^{\ast _{f}}$ is a $\sqrt{N}$--CAN estimator
of $d_{T}^{f}$. Proceeding as previously, replacing $\widehat{d}_{T}$ by $%
\widehat{d}_{T}^{\ast _{f}}$, $d$ by $d_{T}^{f}$ and $E[\widehat{d}_{T}]$ by
$E^{\ast }[\widehat{d}_{T}^{\ast _{f}}]$, we have
\begin{equation}
b_{T}^{\ast }=E^{\ast }[\widehat{d}_{T}^{\ast _{f}}]-d_{T}^{f}  \label{btbar}
\end{equation}%
where $E^{\ast }$ denotes the expectation associated with $(\Omega ^{\ast },%
\mathfrak{F}^{\ast },P^{\ast })$, the probability space induced by the
bootstrap process. Given that $\overline{d}_{T,B}^{\ast _{f}}$ in (\ref%
{dstar_bar}) can be made arbitrarily close to $E^{\ast }[\widehat{d}%
_{T}^{\ast _{f}}]$ by taking $B$ sufficiently large, \eqref{btbar}
represents the estimator of the true finite sample bias induced by the
pre-filtered sieve bootstrap.

The accuracy with which $b_{T}^{\ast }$ estimates $b_{T}$ obviously
underpins the validity of the bootstrap bias correction method. To evaluate
the magnitude of $|b_{T}-b_{T}^{\ast }|$ note that $|\kappa (e^{\imath
\lambda })|^{2}$ (for $\kappa (\cdot )$ as defined in (\ref{Wold})) is a
bounded, even function of $\lambda $, and we have the power series
(McLaurin) expansion $|\kappa (e^{\imath \lambda })|^{2}=c_{0}+\sum_{j\geq
1}c_{j}|\lambda |^{2j}=c_{0}+c_{1}|\lambda |^{2}+o(|\lambda |^{3})$ as $%
|\lambda |\rightarrow 0$. Then, as is shown in \cite%
{hurvich:deo:brodsky:1998} and \cite{giraitis:robinson:2003}, see also %
\citet{andrews:guggenberger:2003} and \cite{andrew:sun:2004},
\begin{equation}
b_{T}=-\beta \frac{2c_{1}}{9c_{0}}\left( \frac{N}{T}\right) ^{2}+o\left(
\frac{N^{2}}{T^{2}}\right)   \label{biasd}
\end{equation}%
for the LPR and SPLW estimators, where $\beta >0$. Similarly, set $\bar{%
\kappa}_{h}(z)=\sum_{j=0}^{\infty }\bar{\kappa}_{h}(j)z^{j}$, where the $%
\bar{\kappa}_{h}(j)$ and $\bar{\phi}_{h}(j)$ are related by the recursion
\begin{equation}
\bar{\phi}_{h}(0)=\bar{\kappa}_{h}(0)=1\,,\;\;\sum_{i=0}^{j}\bar{\kappa}%
_{h}(i)\bar{\phi}_{h}(j-i)=0,\;j=1,2,\ldots \,.  \label{phihinv}
\end{equation}%
By construction $\bar{\kappa}_{h}(z)\bar{\phi}_{h}(z)=1$ for all $|z|\leq 1$
and $\bar{\kappa}_{h}(z)$ yields the $\textit{AR}(h)$ approximation to $\kappa (z)$
implicit in the bootstrap algorithm. Then $|\bar{\kappa}_{h}(e^{\imath
\lambda })|^{2}=|\sum_{j=0}^{h}\bar{\phi}_{h}(j)e^{\imath \lambda j}|^{-2}=%
\bar{c}_{0}+\bar{c}_{1}|\lambda |^{2}+o(|\lambda |^{3})$ as $|\lambda
|\rightarrow 0$ and
\begin{equation}
b_{T}^{\ast }=-\beta \frac{2\bar{c}_{1}}{9\bar{c}_{0}}\left( \frac{N}{T}%
\right) ^{2}+o\left( \frac{N^{2}}{T^{2}}\right) \,.  \label{biasbsd}
\end{equation}

\begin{theorem}
\label{pfsbbias} Suppose that the process $y(t)$ satisfies Assumptions \ref%
{Ass1} and \ref{Ass2}. Assume that $d_{T}^{f}$ is chosen such that $%
|d_{T}^{f}-d|<\delta _{T}$, where $\delta _{T}\log T\rightarrow 0$ as $%
T\rightarrow \infty $, and that an AR$(h)$ approximation is used within the
pre-filtered sieve bootstrap, where $h\leq H_{T}=a(\log T)^{c}$, $a>0$, $%
c<\infty $. Assume also that $b_{T}=E[\widehat{d}_{T}]-d$ and $b_{T}^{\ast
}=E[\widehat{d}_{T}^{\ast _{f}}]-d_{T}^{f}$ are given by expressions %
\eqref{biasd} and \eqref{biasbsd} respectively. Then
\begin{equation*}
\left\vert b_{T}-b_{T}^{\ast }\right\vert =O\left( \max \left\{ h\left(
\frac{\log T}{T}\right) ^{\half-\delta _{T}},\;\delta _{T}h^{-|d|},\;\delta
_{T}\log T\right\} \right) +o\left( \frac{N^{2}}{T^{2}}\right)
\end{equation*}%
almost surely.
\end{theorem}

It is obvious from Theorem \ref{pfsbbias} that $|b_{T}-b_{T}^{\ast }|=o(1)$
a.s., and not surprisingly, that the rate of convergence of $b_{T}^{\ast }$
to $b_{T}$ induced by the PFSB depends on the order of the autoregressive
approximation ($h$) and the proximity of the preliminary filtering value to
the true $d$, that is the value of $\delta _{T}$ implicit in the choice of $%
d_{T}^{f}$. 
Which term in Theorem \ref{pfsbbias} ultimately dominates $%
|b_{T}-b_{T}^{\ast }|$, the $O(h(\log T/T)^{\half-\delta _{T}})$ or the $%
O(\delta _{T}\log T)$, will depend on whether $\delta _{T}\rightarrow 0$
faster or slower than $h/(T\log T)^{1/2}$. Given that the values of the
three Landau \textquotedblleft big-Oh\textquotedblright constants that
appear in Theorem \ref{pfsbbias} have not been quantified, this indicates
that the choice of $h$ and $d_{T}^{f}$ will have an important impact on both
the finite sample and asymptotic behaviour of $b_{T}-b_{T}^{\ast }$.
Selection of $h$ by AIC yields $h\sim K\log T$ $a.s.$ as $T\rightarrow
\infty $, which is asymptotically efficient in the sense of \cite%
{shibata:1980}; see \cite{poskitt:2008} and \citep[][\S 3]{politis:2003}.

Appropriate selection of the pre-filtering value for $d$ is less clear. As
noted earlier, we initially choose as pre-filters the actual estimators that
we are interested in bias correcting, namely the LPR and SPLW estimators and
their analytically bias-reduced variants. Noting that in both cases the
analytic bias reduction involves the inclusion of one or more even powers of
frequency in the respective objective functions (see %
\citealp{andrews:guggenberger:2003} and \citealp{andrew:sun:2004} for
details), we designate the LPR-based and SPLW-based estimators as $\textit{%
LPR}(P)$ and $\textit{SPLW}(P)$ respectively, where $P=0$ indicates the
original \cite{geweke:porter:1983}/\cite{robinson:1995a,robinson:1995b}
estimators; and $P=1,2,\ldots $ indicate the corresponding bias-reduced
variants based on the inclusion of $P$ even powers of the frequencies. The
limiting distributions of the latter are related to those of the former via
a \textquotedblleft variance inflation factor\textquotedblright\ $\psi
_{P}^{2}$; that is,
\begin{equation}
N^{1/2}(\widehat{d}_{T}-d)\overset{\mathcal{D}}{\rightarrow }\mathcal{N}%
\left( 0,\omega ^{2}\psi _{P}^{2}\right) ,  \label{lpr_dist}
\end{equation}%
where $\omega ^{2}={\pi ^{2}}/{24}$ for $\widehat{d}_{T}$ produced via LPR, $%
\omega ^{2}={1}/{4}$ for $\widehat{d}_{T}$ produced via SPLW, $\psi
_{0}^{2}=1$ yields the baseline variance for the uncorrected estimator, and $%
\psi _{P}^{2}$ increases with $P$. In particular, $\psi _{1}^{2}=2.25$, $%
\psi _{2}^{2}=3.52$ and $\psi _{3}^{2}=4.79$.

That each of these estimators can serve as a legitimate pre-filtering value
rests on the following proposition:

\begin{proposition}
\label{lrgdev} Let $d_{T}^{f}$ denote any one of the estimators $\textit{LPR}%
(P)$ or $\textit{SPLW}(P)$, with $P=0,1,2,...$ Then under the conditions of
Theorem \ref{pfsbbias} $|d_{T}^{f}-d|\log T\rightarrow 0$ as $T\rightarrow
\infty $ with probability one.
\end{proposition}

As is made clear in Appendix \ref{proofs}, this proposition follows directly
for the {SPLW}(0) estimator from existing results. However, for the
remaining estimators detailed proofs are required. Furthermore, for a
bootstrap-bias-adjusted version of an initial estimator we have $\widetilde{d%
}_{T}-d=\widehat{d}_{T}-d-\widehat{b}_{T,B}^{\ast _{f}}$, and adding and
subtracting the bootstrap bias before applying the triangle inequality gives
$|\widetilde{d}_{T}-d|\leq |\widehat{d}_{T}-d|+|\widehat{b}_{T,B}^{\ast
_{f}}-b_{T}^{\ast }|+|b_{T}^{\ast }|$. Since the bootstrap estimate of bias
will obey the law of the iterated logarithm (in $B$) we have $|\widehat{b}%
_{T,B}^{\ast _{f}}-b_{T}^{\ast }|=O(\sqrt{\log \log B/B})$ a.s.. Consistency
and asymptotic normality of the estimator also imply that $|b_{T}^{\ast
}|=o(N^{-1/2})$. We therefore conclude that $|\widetilde{d}_{T}-d|\log T\leq
o(1)+\log T\{O(\sqrt{\log \log B/B})+o(N^{-1/2})\}\rightarrow 0$ as $%
T\rightarrow \infty $ for any $B\sim KT^{\beta }$, $\beta >0$, and hence $%
d_{T}^{f}=\widetilde{d}_{T}$ can serve as a valid pre-filtering value in a
subsequent application of the algorithm. This observation prompts the
extension of Section \ref{recursive}, in which successive
bootstrap-bias-adjusted versions of the $\textit{LPR}(P)$ and $\textit{SPLW}%
(P)$ estimators play the role of the preliminary pre-filtering value within
an iterative bias-correction scheme.

\subsection{Asymptotic coverage of bootstrap confidence intervals}

The following theorem links the accuracy of the bias estimation to the
accuracy with which the full sampling distribution of the relevant estimator
is approximated via the bootstrap and, hence, to the coverage accuracy of
the HPD confidence intervals computed using the bootstrap draws.

\begin{theorem}
\label{hpd_coverage} Set
\begin{equation*}
\overline{{\Pr }}^{\ast }\left\{ N^{\half}(\widehat{d}_{T}^{\ast
_{f}}-\overline{d}_{T,B}^{\ast _{f}})<x\right\} =B^{-1}\sum_{b=1}^{B}\mathbf{%
1}\left\{ N^{\half}(\widehat{d}_{T,b}^{\ast _{f}}-\overline{d}_{T,B}^{\ast
_{f}})\leq x\right\}\,.
\end{equation*}
Then under the conditions of Theorem \ref{pfsbbias} it follows that
\begin{equation*}
\sup_{x}\left\vert {\Pr }\left\{ N^{\half}(\widehat{d}_{T}-E[%
\widehat{d}_{T}])<x\right\} -\overline{{\Pr }}^{\ast }\left\{ N^{%
\half}(\widehat{d}_{T}^{\ast _{f}}-\overline{d}_{T,B}^{\ast _{f}})<x\right\}
\right\vert \leq \frac{N^{\half}|b_{T}-b_{T}^{\ast }|}{\upsilon \sqrt{2\pi }}%
+r_{BN}
\end{equation*}%
where the remainder $r_{BN}=N^{1/2}O( \sqrt{\log \log B/B})+o(
N^{5/2}/T^{2}) $.
\end{theorem}

Theorem \ref{hpd_coverage} makes it clear that the distribution of $N^{1/2}(%
\widehat{d}_{T}^{\ast _{f}}-E^{\ast }[\widehat{d}_{T}^{\ast _{f}}])$ will
closely approximate the true finite sampling distribution of $N^{1/2}(%
\widehat{d}_{T}-E[\widehat{d}_{T}])$ provided $N^{1/2}|b_{T}-b_{T}^{\ast }|$
is sufficiently small. Given $N=KT^{v}$ for $\nu \in (2/3,4/5)$, it follows
from Theorem \ref{pfsbbias} that $N^{1/2}|b_{T}-b_{T}^{\ast }|\rightarrow 0$%
, and the accuracy with which the bootstrap-based estimate of the bias
replicates the true bias as $N^{1/2}|b_{T}-b_{T}^{\ast }|$ approaches zero
can be viewed as a representation of the accuracy with which the pre-filtered bootstrap
reproduces the true sampling distribution of the estimator per se.
This implies, in turn, that for $B$ sufficiently large ($B\sim KT^{4/5+\beta
}$, $\beta >0$) HPD $(1-\alpha _{U}-\alpha _{L})100\%$ confidence intervals
constructed from $B$ bootstrap draws will have the correct (asymptotic)
coverage. To wit, use $B$ bootstrap draws to construct the interval $(%
\widehat{d}_{T}-\widehat{q}_{T,B}^{\ast _{f}}(1-\alpha _{U}),\widehat{d}_{T}-%
\widehat{q}_{T,B}^{\ast _{f}}(\alpha _{L}))$, with $\widehat{q}_{T,B}^{\ast
_{f}}(1-\alpha _{U})$ and $\widehat{q}_{T,B}^{\ast _{f}}(\alpha _{L})$
denoting the upper and lower quantiles of the narrowest interval containing $%
(1-\alpha _{U}-\alpha _{L})100\%$ of the bootstrap distribution of the mean
corrected values $\widehat{d}_{T,b}^{\ast _{f}}-\overline{d}_{T,B}^{\ast
_{f}}$, $b=1,\ldots ,B$. Noting from (\ref{bs_bias}) that $\overline{d}%
_{T,B}^{\ast _{f}}=d_{T}^{f}+\widehat{b}_{T,B}^{\ast _{f}}$, we can see that
the intervals so constructed correspond to bootstrap centered percentile
confidence intervals that adjust for bias and accommodate possible asymmetry
about the mean.\footnote{%
See, \textit{inter alia}, \citet[][Chapter 10]{hansen:2014} and
\citet[][Chapter
23]{vaart:1998} for discussions of bootstrap confidence intervals and their
associated properties.}

\section{An Iterative Bias-Correction Procedure\label{recursive}}

Although the bias of $\widetilde{d}_{T}$ in \eqref{bias_adjusted_est} will
be smaller than that of $\widehat{d}_{T}$, the remaining bias $E[\widetilde{d%
}_{T}]-d$ may still be large because the bias in any preliminary value $%
d_{T}^{f}$ can be severe in finite samples, and $\widehat{b}_{T,B}^{\ast
_{f}}$ in \eqref{bs_bias} will, as a consequence, be a biased estimate of
its true counterpart $b_{T}$ in \eqref{bt}. To obtain a more accurate
estimate of $d$ we propose a further refinement to the PFSB-based
bias-correction procedure via a recursive algorithm involving two stochastic
stopping criteria as follows:

\begin{description}
\item[1. Initialization:] \label{iterative} Set $k=0$ and assign desirable
tolerance levels $\tau _{1}=\tau _{1}^{(0)}$ and $\tau _{2}=\tau _{2}^{(0)}$
for the two stopping rules. For the chosen estimator $\widehat{d}_{T}$, set $%
\widetilde{d}_{T}^{(0)}=\widehat{d}_{T}$ (i.e. set $d^{f}=$ $\widehat{d}_{T}$%
).

\item[2. Recursive Calculation:] For the $k^{th}$ iteration set the
preliminary value of $d$, namely $d_{T}^{f},$ to $\widetilde{d}_{T}^{(k)}$
and perform the second and third steps of the bias-correction procedure of
Section \ref{bc} with $\widehat{d}_{T}$ therein replaced by $\widetilde{d}%
_{T}^{(k)}$, producing, in an obvious notation, $\widetilde{d}_{T}^{(k+1)}=%
\widetilde{d}_{T}^{(k)}-\widetilde{b}_{T,B}^{\ast _{f}(k)}$.

\item[3. Stopping Rules:] If $\left\vert \widetilde{d}_{T}^{(k+1)}-%
\widetilde{d}_{T}^{(k)}\right\vert >\tau _{1}$ and $\left\vert \widetilde{d}%
_{T}^{(0)}-\widetilde{d}_{T}^{(k)}-\widetilde{b}_{T,B}^{\ast
_{f}(k)}\right\vert >\tau _{2}$ set $k=k+1$, update the tolerance levels $%
\tau _{1}=\tau _{1}^{(k)}$ and $\tau _{2}=\tau _{2}^{(k)}$, and repeat Step
2. Otherwise set $\widetilde{d}_{T}=\widetilde{d}_{T}^{(k)}$ and stop.
\end{description}

The rationale behind the recursions is as follows: since the estimator $%
d^{f}=\widehat{d}_{T}$ is biased, $\widehat{b}_{T,B}^{\ast _{f}}$ will on
average be a biased estimate of $b_{T}$, and the bias-adjusted estimate $%
\widetilde{d}_{T}$ will therefore still contain some bias. Replacing the
initial values $\widehat{d}_{T}=\widetilde{d}_{T}^{(0)}$ and $\widehat{b}%
_{T,B}^{\ast _{f}}=\widetilde{b}_{T,B}^{\ast _{f}(0)}$ by $\widetilde{d}%
_{T}^{(1)}$ and $\widetilde{b}_{T,B}^{\ast _{f}(1)}$, and (for general $k$) $%
\widetilde{d}_{T}^{(k-1)}$ and $\widetilde{b}_{T,B}^{\ast _{f}(k-1)}$ by $%
\widetilde{d}_{T}^{(k)}$ and $\widetilde{b}_{T,B}^{\ast _{f}(k)}$, and so
on, produces more accurate estimates and bias assessments. Being based upon
more accurate estimators, the updated estimate $\widetilde{d}_{T}^{(k)}$
would be expected to be closer to the true value of $d$. The procedure is
iterated until no meaningful gain in accuracy is achieved. Details of the
two stochastic criteria used to determine when sufficient accuracy has been
attained are given in Appendix \ref{ssr}. Some further comment on stopping
rules is also included in the section following.

\section{Simulation Exercise\label{sim}}

\subsection{Simulation design\label{design}}

In this section we illustrate the performance of the
bootstrap-bias-corrected estimators via a Monte Carlo experiment. Following
\cite{andrews:guggenberger:2003} we simulate data from an $\textit{ARFIMA}%
(1,d,0)$ process,
\begin{equation}
(1-L)^{d}\Phi (z)y(t)=\varepsilon (t)\,,\ 0\leq d<0.5\,,  \label{arfima}
\end{equation}%
where $\Phi (z)=1-\phi z$ is the operator for a stationary $\textit{AR}(1)$ component
and $\varepsilon (t)$ is zero mean white noise, assumed initially to be
Gaussian. The choice of this model is motivated, in part, by earlier work
that highlights the distinct finite sample bias of the LPR estimator of $d$
in this setting, when the value of $\phi $ is positive and large %
\citep[See][]{agiakloglou:newbold:wohar:1993}. Indeed, \cite%
{andrews:guggenberger:2003} document substantial remaining bias in the
bias-corrected version of the LPR estimator in the presence of a large
autoregressive parameter. The impetus for applying bootstrap-based bias
correction to the various estimators is accordingly particularly strong in
this setting.

The process in (\ref{arfima}) is simulated $R=1000$ times for $%
d=0.0,0.2,0.3,0.4$; $\phi =0.3,0.6,0.9$; and sample sizes $T=100$ and $500$
via Levinson recursion applied to the autocovariance function of the desired
$\textit{ARFIMA}(p,d,q)$ process and the generated pseudo-random $%
\varepsilon (t)$ \citep[see, for
instance,][\S5.2]{brockwell:davis:1991}. The ACF for given $T$, $\phi $, $%
\theta $, and $d$ is calculated using Sowell's \citeyearpar{sowell:1992}
algorithm as modified by \cite{doornik:ooms:2003}.

The estimators to which we apply the iterative bias-correction procedure of
Section \ref{recursive} are $\textit{LPR}(P)$ and $\textit{SPLW}(P)$, $%
P=0,1,2$, implemented with bandwidth $N=T^{0.7}$. The value of $N$ accords
with common practice, with the exponent falling within the $(2/3,4/5)$ range.%
\footnote{%
A lower bound of $N\sim KT^{2/3}$ reflects the fact that unless $N$
increases sufficiently quickly with $T$ terms of order $O(\log ^{3}N/N)$ in
the expansions of $b_{T}$ and $b_{T}^{\ast }$ compete with the terms in %
\eqref{biasd} and \eqref{biasbsd}; and the upper bound reflects that the
estimators are known to be rate optimal when $N\sim KT^{4/5}$ in the
uncorrected case \citep{giraitis:robinson:samarov:1997} and $N\sim
KT^{(4+4P)/(5+4P)}$ in the corrected case %
\citep{andrews:guggenberger:2003,andrew:sun:2004}, although asymptotic
normality of the estimators requires that $N=o(T^{4/5})$.}
The order ($h$) of the autoregressive approximation underlying the sieve
component of the bootstrap algorithm is chosen via AIC, and Burg's algorithm
is used to estimate the autoregressive parameters. The number of bootstrap
realizations is $B=1000$.

We compute the empirical bias and MSE for the original estimators prior to
bootstrap-based bias correction (i.e., $\textit{LPR}(P)$ and $\textit{SPLW}%
(P)$, $P=0,1,2$), and for the bootstrap-bias-corrected versions thereof. The
latter are produced through formal application of the stochastic stopping
rules described in Appendix \ref{ssr}, augmented by a deterministic
criterion, whereby the iterative scheme ceases if $\widetilde{d}%
_{T}^{(k+1)}<-1$ or $\geq 1.5$ and the estimator $\widetilde{d}_{T}^{(k)}$
retained as the final choice. We also report bias and MSE results for the
bootstrap-bias-corrected estimators based on the first two iterations of the
iterative procedure. This comparison of the sampling properties of
estimators with varying degrees of analytical bias correction with those of
estimators that exploit the bootstrap bias adjustment, allows us to
investigate, firstly, the efficacy of using the bootstrap method rather than
an analytical method to bias adjust; and, secondly, the possibility of
obtaining additional improvement by bias-correcting (via the bootstrap) an
estimator that has already been bias-adjusted via analytical means. Finally,
as a reference for the magnitude of the bias and MSE of the various raw and
bias-adjusted semiparametric estimators, we record\textbf{\ }the
corresponding statistics\ for the correctly specified MLE.\footnote{%
Numerical evidence presented in \cite{nielsen:frederiksen:2005} suggests
that semiparametric estimators can be competitive with correctly specified
parametric methods. Comparison of the performance of the semiparametric
estimators with that of the correctly specified, and hence asymptotically
optimal, MLE is therefore of interest.}

We also compute the empirical coverage and length of nominal 95\% HPD
intervals obtained by applying the bootstrap procedure to the estimators $%
\textit{LPR}(P)$ and $\textit{SPLW}(P)$, $P=0,1,2$. For each of the $R$
Monte Carlo replications the intervals are constructed as described in
Section \ref{theory}, each in turn based on $B$ bootstrap draws, and the
empirical coverage is calculated as the proportion of times (in $R$
replications) that each interval includes the true value of $d$. The
empirical length of the intervals is recorded as their average length across
the $R$ replications. These coverage and length statistics are compared with
the empirical coverage and (constant) length of 95\% confidence intervals
constructed from the appropriate asymptotic distributions in (\ref{lpr_dist}%
). Results did not vary markedly with $d$, and hence are presented averaged
over the four values of $d$ considered.\footnote{%
Analogous results for nominal 90\% HPD intervals were found to be
qualitatively similar in all cases and, hence, are not reported or
explicitly discussed.}

In line with the assumption of Gaussianity, thus far we have supposed that
the bootstrap innovations generated in Step 2(c) of the PFSB algorithm are $%
i.i.d.$ $\mathcal{N}(0,\bar{\sigma}_{h}^{2})$. Such bootstrap realizations
are said to be generated via a parametric bootstrap. Nonparametric bootstrap
innovations can be generated using the following modification of Step 2(c):

\begin{description}
\item[2(c$^{\prime }$)] Let $\varepsilon _{h}^{+}(t)$, $t=1,\ldots ,T$,
denote a simple random sample of \textit{i.i.d.} values drawn from $U_{%
\tilde{\varepsilon}_{h},T}(e)=T^{-1}\sum_{t=1}^{T}\mathbf{1}\{\tilde{%
\varepsilon}_{h}(t)\leq e\}$, the probability distribution function that
places a probability mass of $1/T$ at each of $\tilde{\varepsilon}_{h}(t)$, $%
t=1,\ldots ,T$. Set $\varepsilon _{h}^{\ast }(t)=\bar{\sigma}_{h}\varepsilon
_{h}^{+}(t)$, $t=1,\ldots ,T$.
\end{description}

\noindent The innovations generated by the nonparametric bootstrap are $%
i.i.d.$ $(0,\bar{\sigma}_{h}^{2})$ by construction, and when $y(t)$ is
Gaussian we can expect $\varepsilon _{h}^{\ast }(t)$, $t=1,\ldots ,T$,
and hence $y^{\ast _{f}}(t)$, to be approximately Gaussian. This suggests
that replacing the innovations generated in PFSB-2(c) by those generated in
2(c$^{\prime }$) should not produce outcomes that are substantially
different, and we document this by presenting some selected results in which
Gaussianity is retained for the data generating process, but nonparametric
bootstrap innovations are generated as per 2(c$^{\prime }$) above. PFSB-2(c$%
^{\prime }$) also caters for the possibility that $y(t)$ is a linear process
with innovations that do not satisfy Assumption \ref{Ass2}, and so allows us
to examine the robustness of our results to violations of the assumption of
Gaussianity. Accordingly, we report some selected results obtained using the
nonparametric pre-filtered bootstrap with a Student $t$ distribution adopted
for $\varepsilon (t)$.

In summary, Tables \ref{table-lpr-100}-\ref{table-hpd95} record results
based on the parametric version of the bootstrap, with Gaussian errors
adopted in (\ref{arfima}); Tables \ref{table-lpr-np} and \ref{table-lpw-np}
record selected results based on the replacement of Step 2(c) in the PFSB
algorithm with the nonparametric 2(c$^{\prime }$) with Gaussian errors
retained; while the results recorded in Tables \ref{table-lpr-st} and \ref%
{table-lpw-st} use the nonparametric version of the bootstrap and assume
Student $t$ innovations. Note that for brevity specific results for $d=0.3$\
are omitted from Tables \ref{table-lpr-100}-\ref{table-lpw-500}, while the
results in the subsequent tables are reported after averaging over all four
values of $d$, including $d=0.3$. To shed some light on the effect of
misspecification on the relative performance of the semi-parametric
estimators (with the nonparametric version of the bootstrap used) and the
MLE, the results for the Gaussian MLE under Student $t$ innovations are
summarized in the final column of Tables \ref{table-lpr-st} and \ref%
{table-lpw-st}. To aid interpretation, the MLE results under Gaussian
innovations (as recorded in Tables \ref{table-lpr-100}-\ref{table-lpw-500})
are also averaged across the four $d$ values and reported in the final
columns of Tables \ref{table-lpr-np} and \ref{table-lpw-np}. All tables are
included in Appendix \ref{tbls}, with the most favorable result (within the
semiparametric set) for each design setting highlighted. The columns headed
`SSR' in certain of the tables record the results based on the stochastic
stopping rules discussed in Appendix \ref{ssr} and modified as described
above.

\subsection{Simulation results: LPR}

Tables \ref{table-lpr-100} and \ref{table-lpr-500} record (for $T=100$ and $%
T=500$ respectively) the bias and MSE results for all estimators based on
the LPR method, using the parametric version of the bootstrap, and with
Gaussian errors adopted in (\ref{arfima}). All results pertaining to the use
of the bootstrap to bias adjust $\textit{LPR}(P)$ are indicated by appending
`$\textit{-BBA}(K)$' to the LPR acronym, where $K$ is the number of times
the bootstrap-bias-correction procedure is applied.\footnote{%
That is, $K=1$ refers to the single application of the bias-correction
procedure without iteration, while $K=2$ (3, etc.) corresponds to $k=1$ (2,
etc.) iterations in the iterative version of the algorithm.} The key message
is that the bootstrap technique \emph{does }reduce bias, but with the most
substantial gains to be had by using the bootstrap algorithm to bias adjust
an estimator that has already been bias reduced analytically. For example,
for $T=100$, and for eight of the nine cases, the smallest bias is produced
by bias adjusting (via the bootstrap) either the ${LPR}(1)$ or ${LPR}(2)$
estimator at least once. For $T=500$, the same qualitative result holds,
with $\textit{LPR}(2)\textit{-BBA}(1)$ being the least biased estimator
overall. Importantly, for $T=500$ at least, the reduction in bias is so
substantial that this estimator also has the lowest MSE of all estimators
(including those not bias adjusted) for $\phi =0.9$ and all values of $d$.
Moreover, even when the bootstrap bias adjustment does cause the (expected)
increase in MSE, it is not excessive.\footnote{%
Note that \emph{all} versions of the LPR estimator, including the
bootstrap-bias-corrected versions, are very biased when $\phi =0.9$. This
confirms (as documented in the literature \textit{op.~cit.}) that
semiparametric estimators experience problems in this part of the parameter
space. The use of our procedure does, however, \emph{reduce} the bias,
indicating that even in this worst case scenario appreciable gains can be
made.}

The results recorded in Table \ref{table-lpr-100} indicate that the stopping
rules are useful for the smaller sample size, producing estimators with the
smallest bias in seven cases. For ${LPR}(1)$, for which results for both $%
K=1 $ and $2$ are recorded, the MSE for the SSR method is seen to fall
in-between the corresponding figures based on these fixed numbers of
applications of the bootstrap in virtually all instances. The results in
Table \ref{table-lpr-500}, however, demonstrate that for $T=500$ a fixed
number of bootstrap-based bias adjustments is preferable overall, with the
SSR method yielding less gains. Hence, and with due consideration taken of
the limitations of the experimental design, we can conclude that although a
stopping rule tailors the number of iterations to the realization at hand,
its use does not appear to \emph{guarantee} an improvement in overall
performance compared to using a fixed number of iterations, at least when
the sample is reasonably large. Note that the finding that the bias results
for the bootstrap-bias-adjusted estimators are superior overall also applies
to the results based on the nonparametric version of the PFSB algorithm, and
under both the Gaussian and Student $t$ errors, as can be seen by the
location of the highlighted figures in Tables \ref{table-lpr-np} and \ref%
{table-lpr-st} respectively.

Another result of interest pertains to the relationship between the overall
accuracy of the bootstrap-bias-corrected estimators (as measured by MSE) and
that of the comparable analytically-adjusted estimators. For instance, $K$
bias adjustments of an $\textit{LPR}(P)$ estimator via the bootstrap can --
for some designs -- yield an estimator with a smaller MSE than does the
equivalent number of analytical bias adjustments. In certain cases this
reduction in MSE goes hand in hand with a smaller bias.\footnote{%
For example, for $T=100$, $d=0.4$ and $\phi =0.3:$ the ${LPR}(1)${-}${BBA}%
(1) $ estimator has an MSE that is notably less than that of the `matching' $%
{LPR}(2)$ estimator, at the cost of only a small increase in bias. For the
same parameter design, but with $T=500$, the first estimator has both
smaller bias and smaller MSE than the second.} We return to this point in
Section \ref{retr}. With regard to overall accuracy, our results are also in
accord with the findings of \cite{nielsen:frederiksen:2005} in that, in the
absence of persistent short-run dynamics (i.e. $\phi \neq 0.9$), the
bootstrap-bias-corrected semiparametric estimators often exhibit significant
bias reduction relative to the correctly specified MLE. That this
improvement is at the expense of an increase in MSE relative to the MLE is
perhaps not surprising given that the correctly specified MLE is
(asymptotically) optimal. From a comparison of the results recorded in the
final column of Tables \ref{table-lpr-np} and \ref{table-lpr-st} it is
evident that misspecification of the innovations as Student $t$ has little
impact on the performance of the Gaussian MLE. Hence, the qualitative nature
of the comment made above regarding the relative performance of the
semiparametric and parametric estimators continues to obtain.

Finally, the most notable characteristic of the HPD results in Panel A of
Table \ref{table-hpd95} (produced under the parametric algorithm) is the
improvement in coverage yielded by the bootstrap technique,\textbf{\ }%
relative to that yielded by the relevant asymptotic distribution. In
particular,\textbf{\ }use of the PFSB distributions produces intervals with
close to correct coverage for the\textbf{\ }low and moderate values of $\phi
$, and for the estimators based on\textbf{\ }$P\geq 1.$ Unsurprisingly,
while for any particular $\textit{LPR}(P)$ estimator (i.e. for any given
value of $P$) the improvement in coverage accuracy is accompanied by an
increase in interval width, this decrease in precision is not excessive. In
both the bootstrap and asymptotic cases, an increase in $P$ tends to lead to
an improvement in coverage accuracy, but at the expense (as would be
expected) of an increase in interval width, due to the larger variance of
the underlying estimator.

\subsection{Simulation results: SPLW\label{splw_results}}

Tables \ref{table-lpw-100} and \ref{table-lpw-500} record (for $T=100$ and $%
500$ respectively) the bias and MSE results for all estimators based on the
SPLW method (with the postfix `$\textit{-BBA}(K)$' used to indicate
bootstrap bias adjustment as described above). As with the LPR-based
estimators, the bootstrap-based bias adjustment yields the largest bias
reductions overall, but (in all but one case) only when applied to an SPLW
estimator that has already been analytically bias adjusted. Most notably,
the SSR method, specifically as applied to {SPLW}(2), yields the best bias
reductions overall, and for both sample sizes; although as was the case for
the LPR results, it does not guarantee an improvement in performance over
using a fixed number of iterations.

The biases of all SPLW-based estimators are broadly similar in magnitude to
those of the comparable LPR-based estimators, and as with the LPR-based
estimators, the reduction in bias produced by the bootstrap technique (in
certain cases) is not\ obtained at the expense of MSE. Once again, the
qualitative results regarding bias adjustment still hold when the
nonparametric form of the bootstrap is used, and when Student $t$ errors
feature in the true DGP, rather than Gaussian errors, as seen from the
results recorded in Tables \ref{table-lpw-np} and \ref{table-lpw-st}
respectively. The MSE results demonstrate that the increase in precision
sometimes yielded by the bootstrap over and above a comparable number of
analytical adjustments, in the case of LPR estimator, continues to obtain in
the SPLW case, also at times allied with a reduction in bias.\textbf{\ }As
with the LPR results, the best performing bootstrap-bias-corrected
SPLW-based estimators are often substantially less biased than the MLE (both
correctly and incorrectly specified), for $\phi \neq 0.9$ at least, but at
the expense of MSE as expected. Once again, the bootstrap yields HPD
intervals with notably better coverage than those associated with the
asymptotic distribution, but at some cost in precision.

\subsection{Retrospective\label{retr}}

Our simulation results raise the question of how the sieve bootstrap as
implemented in the PFSB algorithm is able to bias correct the basic LPR and
SPLW estimators, and the analytically-bias-adjusted versions thereof,
without necessarily incurring a substantial, if any, loss in overall
precision.

By way of explanation for this phenomenon, consider the LPR estimator. This
estimator is commonly motivated by observing that
\begin{equation}
\frac{I_{T}(\lambda_j )2\pi |1-e^{-\imath \lambda_j }|^{2d}}{\sigma
^{2}|\kappa (e^{\imath \lambda_j })|^{2}}\overset{\mathcal{D}}{\rightarrow }%
V_j\,,  \label{perio}
\end{equation}%
wherein $I_{T}(\lambda )$ denotes the periodogram and $V_j$ is distributed
exponentially when $d=0$, and as an unequally weighted sum of independent $%
\chi^2(1)$ random variables when $d\neq 0$
\citep[][Theorem
6]{hurvich:beltrao:1993}. Taking logarithms in \eqref{perio} and using the
approximation $|1-e^{-\imath \lambda }|^{2d}=|\lambda |^{2d}(1+o(1))$ as $%
\lambda\rightarrow 0$ leads to the linear regression model
\begin{equation}
\log (I_{T}(\lambda _{j}))=\alpha _{0}-2d\log (\lambda _{j})+\eta _{j},
\label{lpr1}
\end{equation}%
where $E[\eta _{j}]=0$ and the intercept $\alpha _{0}$ is presumed to
capture the effects of the adjustments
\begin{align}
a_{j}& =\log |\kappa (1)|^{2}+\log \left( \frac{|\kappa (e^{\imath \lambda
_{j}})|^{2}}{|\kappa (1)|^{2}}\right) -d\log \left( \frac{|1-e^{-\imath
\lambda _{j}}|^{2}}{\lambda _{j}^{2}}\right) -C_j  \label{lpr2} \\
& =\log |\kappa (1)|^{2}-C_j+O(N^{2}/T^{2})\quad \mbox{for all}\quad 1\leq
j\leq N~,  \label{lpr2b}
\end{align}%
where the mean correction term $C_j\leq 0.577216$ (Euler's constant).%
\footnote{%
The expression in \eqref{lpr2b} follows as a consequence of the fact that $%
\log (|\kappa (e^{\imath \lambda })|^{2}/|\kappa (1)|^{2})=\log
(1+(c_{1}/c_{0})|\lambda |^{2}+o(|\lambda |^{3}))$ and $\log (|1-e^{-\imath
\lambda }|^{2}/\lambda ^{2})=\log (1-(1/12)|\lambda |^{2}+o(|\lambda |^{3}))$
as $\lambda \rightarrow 0$.} The presumption that $\alpha _{0}$ absorbs the
effects of the adjustment terms assumes the $a_{j}$ approach a constant as $%
T $ increases and \citep[see the discussion in][]{hurvich:beltrao:1993} it
is the failure of this assumption that can be a source of bias.

The analytical correction replaces the simple regression in \eqref{lpr1} by
the multiple regression
\begin{equation}
\log (I_{T}(\lambda _{j}))=\sum_{p=0}^{P}\alpha _{p}\lambda _{j}^{2p}-2d\log
(\lambda _{j})+\eta _{j}\,,  \label{lpr3}
\end{equation}%
the rationale being that the term $\sum_{p=0}^{P}\alpha _{p}\lambda
_{j}^{2p} $ provides a better approximation to the Maclaurin series
expansion of the right hand side of \eqref{lpr2} than supposing $a_{j}$ is
constant in a neighbourhood of zero. The introduction of $\lambda _{j}^{2p}$%
, $p=1,\ldots ,P$, in \eqref{lpr3} reduces the bias in the estimate of $d$,
but it is also the presence of these additional regressors that causes the
variance inflation seen in \eqref{lpr_dist}.

The pre-filtered bootstrap, on the other hand, takes the specification of
the regression in \eqref{lpr1} or \eqref{lpr3} as given and adjusts the
estimator by mimicking the sampling behaviour of the regressand.
Recall that $I_{T}(\lambda )=(2\pi )^{-1}\sum_{r=1-T}^{T-1}\widehat{\gamma }%
(r)e^{\imath \lambda r}$ where $\widehat{\gamma }(r)=\widehat{\gamma }(-r)$,
$r=0,1,\ldots ,T-1$, denotes the sample autocovariance function. \cite%
{hosking:1996} shows that even in moderate to large samples the $\widehat{%
\gamma }(r)$ have substantial negative bias relative to the true
autocovariances when $d$ is large. 
The bootstrap works by reducing the value of the fractional integration
parameter in the \textquotedblleft data\textquotedblright\ to which the
sieve bootstrap is applied, via the pre-filtering procedure. This reduces
the aforementioned bias. The reduction in $d$ also increases the proximity
of the $C_{j}$ in \eqref{lpr2b} to Euler's constant and renders the $\eta
_{j}$ in \eqref{lpr1} and \eqref{lpr3} closer to centered Gumbel random
variables with variance $\pi ^{2}/6$.
Whether it is applied to \eqref{lpr1} or \eqref{lpr3}, the bootstrap is
thereby able to attack the problem of bias in the estimation of $d$ without
compromising (indeed, reinforcing) the fundamental result assumed to
underlie log periodogram regression and determine the estimators' variance,
namely, the pivotal nature of the ratio in \eqref{perio}.

Although the underlying reasoning is somewhat heuristic, the previous
arguments provide a straightforward explanation of how the pre-filtered
bootstrap is able to exhibit the type of creditable performance observed in
simulation when it is used to bias correct the LPR estimator. Similar
arguments can also be employed to explain the performance characteristics
seen when the algorithm is applied to the SPLW estimator.

\section{Discussion}

This paper has developed a bootstrap method for bias correcting
semiparametric estimators of the long memory parameter in fractionally
integrated processes. The method involves applying the sieve bootstrap to
data pre-filtered by a preliminary semiparametric estimate of the long
memory parameter. In addition to providing theoretical (asymptotic)
justification for using the bootstrap techniques to bias correct, we
document the results of simulation experiments, in which the finite sample
performance of the bootstrap-bias-corrected estimators is compared with that
of both unadjusted estimators and estimators adjusted via analytical means.
The numerical results are very encouraging, and suggest that the bootstrap
bias correction can yield accurate inference about long memory dynamics in
the types of samples that are encountered in practice -- most notably when
applied to estimators for which some preliminary analytical bias reduction
has been used. The bootstrap method is also shown to yield (asymptotically)
valid confidence intervals that formally adjust for bias, with the empirical
coverage of the bootstrap intervals being much closer to the nominal level
than is the coverage of intervals based on the asymptotic distributions of
the relevant semiparametric estimators.

Our discourse has focused on stationary long memory fractional processes,
but as noted previously the pre-filtered sieve bootstrap algorithm does not
restrict either the estimator that is to be bias corrected or the pre-filter
to lie in the interval $[0,0.5)$. The broader applicability of the
pre-filtered sieve bootstrap to more general processes, and to estimators
and pre-filters capable of handling this generality, is therefore only
contingent on establishing its theoretical validity, and as is apparent from
Theorems \ref{pfsbbias} and \ref{hpd_coverage} this hinges on showing that
the pre-filtering value $d_{T}^{f}$ is such that $|d_{T}^{f}-d|\log T=o(1)$ $%
a.s.$. If $N^{1/2}(d_{T}^{f}-d)$ were\emph{\ exactly} $N(0,\upsilon )$ then
it would follow from the tail area properties of the normal distribution
that this condition would be satisfied. Unfortunately, approximate
Gaussianity associated with a pre-filtering value $d_{T}^{f}$ being a $\sqrt{%
N}$-CAN estimator of $d$ -- as has been established for the more general
estimators of \citet{tanaka:1999} and \citet{shimotsu:phillips:2005} -- is
not sufficient to derive the required result because departures of $%
(d_{T}^{f}-d)$ from zero that are inconsequential for weak convergence need
not be so for large-deviation probabilities. Nevertheless, the necessary
large-deviation property can be derived on a case by case basis, as we have
demonstrated for the LPR and SPLW estimators for the stationary case. It
therefore seems reasonable to suppose that arguments similar to those
employed in the proof of Proposition \ref{lrgdev} can be used to show that
the condition $|d_{T}^{f}-d|\log T=o(1)$ $a.s.$ will also be satisfied by
the aforementioned more general estimators, and under more general data
generating processes, but demonstration of this is beyond the scope of the
current paper. We hope to extend our results on the pre-filtered sieve
bootstrap to the non-stationary case in future research.

\appendix


\section{Proofs\label{proofs}}

\paragraph{Proof of Theorem \protect\ref{pfsbbias}:}

Simple algebraic manipulation applied to \eqref{biasd} and \eqref{biasbsd}
gives us the following bound
\begin{align}
|b_{T}-b_{T}^{\ast }|& =\left\vert \frac{\bar{c}_{1}}{\bar{c}_{0}}-\frac{%
c_{1}}{c_{0}}\right\vert O\left( \frac{N^{2}}{T^{2}}\right) +o\left( \frac{%
N^{2}}{T^{2}}\right)  \notag \\
& \leq\left( \left\vert \frac{c_{1}(\bar{c}_{0}-c_{0})}{c_{0}\bar{c}_{0}}%
\right\vert +\left\vert \frac{(\bar{c}_{1}-c_{1})}{\bar{c}_{0}}\right\vert
\right) O\left( \frac{N^{2}}{T^{2}}\right) +o\left( \frac{N^{2}}{T^{2}}%
\right)\,.  \label{bias_diff}
\end{align}
From the first term on the right-hand-side of (\ref{bias_diff}) it can be
seen that the order of magnitude of $|b_{T}-b_{T}^{\ast }|$ depends on that
of $(\bar{c}_{0}-c_{0})$ and $(\bar{c}_{1}-c_{1})$, and that the larger the
bandwidth $N$, with the attendant increase in bias, the closer the
approximation $|\bar{\kappa}_{h}(e^{\imath \lambda })|^{2}$ invoked by the
algorithm needs to be to the true $|\kappa(e^{\imath \lambda })|^{2}$, for $|%
\bar{\kappa}_{h}(e^{\imath \lambda })|^{2}-|\kappa(e^{\imath \lambda
})|^{2}=(\bar{c}_{0}-c_{0})+(\bar{c}_{1}-c_{1})|\lambda |^{2}+o(|\lambda
|^{3})$ in a neighbourhood of the origin. Since, trivially, $2\beta
N^{2}/9T^{2}
=O(N^{2}/T^{2})$, in order to establish the theorem it is sufficient to show
that $|\bar{c}_{0}-c_{0}|$ and $|\bar{c}_{1}-c_{1}|$ are of order $O\left(
T^{2}M_{T}/N^{2}\right) $ or smaller where $M_{T}=\max \{h(\frac{\log T}{T}%
)^{\half-\delta _{T}},\delta _{T}h^{-|d|},\delta _{T}\log T\}$. The
magnitude of both $(\bar{c}_{0}-c_{0})$ and $(\bar{c}_{1}-c_{1})$ can be
derived from the following lemma.

\begin{lemma}
\label{kk} Assume that the conditions of Theorem \ref{pfsbbias} hold. Let $%
\phi _{h}^{f}(z)=\sum_{j=0}^{h}\phi _{h}^{f}(j)z^{j}$ where $\phi
_{h}^{f}(1),\ldots ,\phi _{h}^{f}(h)$ denote the coefficients in the $%
\textit{MMSEP}(h)$ of the process $w^{f}(t)$ in (\ref{wft}), with $%
d^{f}=d_{T}^{f},$ and let $\sigma _{h}^{f2}$ denote the minimum mean squared
error. Set $\kappa ^{f}(z)=\kappa (z)/(1-z)^{d-d_{T}^{f}}$ and define $%
\kappa _{h}^{f}(z)=\{\phi _{h}^{f}(z)\}^{-1}$ by replacing the coefficients
of $\bar{\phi}_{h}(z)$ by those of $\phi _{h}^{f}(z)$ in the recursions in
equation \eqref{phihinv}.

Then
\begin{equation*}
\lim_{T\rightarrow \infty }\left\vert |\bar{\kappa}_{h}(e^{\imath \lambda
})|^{2}-|\kappa (e^{\imath \lambda })|^{2}\right\vert \leq \nu _{1,T}+\nu
_{2,T}+\nu _{3,T}
\end{equation*}%
where for all $\lambda \in \lbrack 2\pi /T,2\pi N/T]$
\begin{alignat*}{3}
\nu _{1,T}& =\left\vert |\bar{\kappa}_{h}(e^{\imath \lambda })|^{2}-|\kappa
_{h}^{f}(e^{\imath \lambda })|^{2}\right\vert & &= O(h(\log T/T)^{\half%
-\delta _{T}}) &  \\
\nu _{2,T}& =\left\vert |\kappa _{h}^{f}(e^{\imath \lambda })|^{2}-|\kappa
^{f}(e^{\imath \lambda })|^{2}\right\vert & &= O(\delta _{T}h^{-|d|})\quad
\text{and} &  \\
\nu _{3,T}& =\left\vert |\kappa ^{f}(e^{\imath \lambda })|^{2}-|\kappa
(e^{\imath \lambda })|^{2}\right\vert & &= O(\delta _{T}\log T)\, &
\end{alignat*}%
with probability one.
\end{lemma}


\paragraph{Proof of Lemma \protect\ref{kk}:}

Addition and subtraction, and straightforward manipulation, yields
\begin{align}
|\bar{\kappa}_{h}(e^{\imath \lambda })|^{2}-|\kappa (e^{\imath \lambda
})|^{2}& =\left( |\bar{\kappa}_{h}(e^{\imath \lambda })|^{2}-|\kappa
_{h}^{f}(e^{\imath \lambda })|^{2}\right) +\left( |\kappa _{h}^{f}(e^{\imath
\lambda })|^{2}-|\kappa ^{f}(e^{\imath \lambda })|^{2}\right)  \notag \\
& \qquad +\left( |\kappa ^{f}(e^{\imath \lambda })|^{2}-|\kappa (e^{\imath
\lambda })|^{2}\right) \,.  \label{kkbar}
\end{align}

Consider the first term in \eqref{kkbar}, $|\bar{\kappa}_{h}(e^{\imath
\lambda })|^{2}-|\kappa _{h}^{f}(e^{\imath \lambda })|^{2}$. By definition
\begin{equation*}
\bar{\kappa}_{h}(z)-\kappa _{h}^{f}(z)=\frac{\phi _{h}^{f}(z)-\bar{\phi}%
_{h}(z)}{\bar{\phi}_{h}(z)\phi _{h}^{f}(z)}\,,
\end{equation*}%
and since $\bar{\phi}_{h}(z)\neq 0$ and $\phi _{h}^{f}(z)\neq 0$, $|z|\leq 1$%
, there exists an $\epsilon >0$ such that
\begin{align*}
\left\vert \bar{\kappa}_{h}(z)-\kappa _{h}^{f}(z)\right\vert & \leq \epsilon
^{-2}\left\vert \phi _{h}^{f}(z)-\bar{\phi}_{h}(z)\right\vert \\
& \leq \epsilon ^{-2}\sum_{j=0}^{h}\left\vert \phi _{h}^{f}(j)-\bar{\phi}%
_{h}(j)\right\vert \quad \text{for all}\quad |z|\leq 1\,.
\end{align*}%
But
\begin{align*}
\sum_{j=0}^{h}\left\vert \phi _{h}^{f}(j)-\bar{\phi}_{h}(j)\right\vert &
\leq \left( h\sum_{j=0}^{h}|\phi _{h}^{f}(j)-\bar{\phi}_{h}(j)|^{2}\right) ^{%
\half} \\
& =O\left( h\left( \frac{\log T}{T}\right) ^{\half(1-2\max
\{0,d-d_{T}^{f}\})}\right) \\
& =O\left( h\left( \frac{\log T}{T}\right) ^{\half-\delta _{T}}\right) \quad
a.s.
\end{align*}%
by an application of \citet[][Theorem 5]{poskitt:2007} and the fact that $%
|d_{T}^{f}-d|<\delta _{T}$ by assumption. It follows that $|\bar{\kappa}%
_{h}(e^{\imath \lambda })-\kappa _{h}^{f}(e^{\imath \lambda })|=O(h(\log
T/T)^{\half-\delta _{T}})$ $a.s.$ uniformly in $\lambda $, and hence that $%
\left\vert |\bar{\kappa}_{h}(e^{\imath \lambda })|^{2}-|\kappa
_{h}^{f}(e^{\imath \lambda })|^{2}\right\vert =O(h(\log T/T)^{\half-\delta
_{T}})$ $a.s.$ uniformly in $\lambda $. We can therefore interchange limit
operations \citep[][Theorem
13.3]{apostol:1960} to give
\begin{equation*}
\lim_{T\rightarrow \infty }\lim_{\lambda \rightarrow 0}\left\vert |\bar{%
\kappa}_{h}(e^{\imath \lambda })|^{2}-|\kappa _{h}^{f}(e^{\imath \lambda
})|^{2}\right\vert =\lim_{\lambda \rightarrow 0}\lim_{T\rightarrow \infty
}\left\vert |\bar{\kappa}_{h}(e^{\imath \lambda })|^{2}-|\kappa
_{h}^{f}(e^{\imath \lambda })|^{2}\right\vert \,,
\end{equation*}%
which implies that $\nu _{1,T}=O(h(\log T/T)^{\half-\delta _{T}})$ $a.s.$
for all $\lambda \in \lbrack 2\pi /T,2\pi N/T]$.

For the second term in \eqref{kkbar}, $|\kappa _{h}^{f}(e^{\imath \lambda
})|^{2}-|\kappa ^{f}(e^{\imath \lambda })|^{2}$, we have
\begin{equation*}
\kappa _{h}^{f}(z)-\kappa ^{f}(z)=\frac{1-\kappa ^{f}(z)\phi _{h}^{f}(z)}{%
\phi _{h}^{f}(z)}\,,
\end{equation*}%
giving us the bound
\begin{equation*}
\left\vert \kappa _{h}^{f}(z)-\kappa ^{f}(z)\right\vert \leq \epsilon
^{-1}\left\vert 1-\kappa ^{f}(z)\phi _{h}^{f}(z)\right\vert \quad \text{for
all}\quad |z|\leq 1\,.
\end{equation*}%
Let $\rho _{h}(z)=\sum_{j\geq 1}\rho _{h}(j)z^{j}=1-\kappa ^{f}(z)\phi
_{h}^{f}(z)$. Then from Parseval's relation
\begin{equation*}
\sum_{j\geq 1}\rho _{h}(j)^{2}=\int_{-\pi }^{\pi }\left\vert 1-\kappa
^{f}(e^{\imath \lambda })\phi _{h}^{f}(e^{\imath \lambda })\right\vert
^{2}d\lambda =2\pi \sigma ^{-2}\left( \sigma _{h}^{f2}-\sigma ^{2}\right)
\end{equation*}%
and from the Levinson--Durbin recursions \citep{levinson:1947,durbin:1960}
we have $\sigma _{h}^{f2}=(1-\phi _{h}^{f}(h)^{2})\sigma _{h-1}^{f2}$.
Substituting sequentially in the recurrence formula $\sigma _{h}^{f2}=\sigma
_{h+1}^{f2}+\phi _{h}^{f}(h)^{2}\sigma _{h}^{f2}$ leads to the series
expansion $\sigma _{h}^{f2}-\sigma ^{2}=\sum_{r=h}^{\infty }\phi
_{r}^{f}(r)^{2}\sigma _{r}^{f2}$, from which we obtain the bound
\begin{equation*}
\sum_{j\geq 1}\rho _{h}(j)^{2}\leq 2\pi \sigma ^{-2}E\left[ w^{f}(t)^{2}%
\right] \sum_{r=h}^{\infty }\phi _{r}^{f}(r)^{2}\,.
\end{equation*}%
But $\phi _{h}^{f}(h)\sim |d_{T}^{f}-d|/h$ as $h\rightarrow \infty $ %
\citep{inoue:2002,inoue:kasahara:2004} and therefore we can infer that
\begin{equation*}
\sum_{j\geq 1}\rho _{h}(j)^{2}\leq \text{const.}\frac{|d_{T}^{f}-d|^{2}}{%
h^{2|d|}}\zeta (2(1-|d|)),
\end{equation*}%
where $\zeta (\cdot )$ denotes the Riemann zeta function. It follows that $%
\lim_{h\rightarrow \infty }\rho _{h}(e^{\imath \lambda })=0$ and that $%
\lim_{T\rightarrow \infty }|\rho _{h}(e^{\imath \lambda })|^{2}=O(\delta
_{T}^{2}h^{-2|d|})$ almost everywhere on $[-\pi ,\pi ]$. Hence we can
conclude that $\lim_{T\rightarrow \infty }\lim_{\lambda \rightarrow
0}\left\vert |\kappa _{h}^{f}(e^{\imath \lambda })|^{2}-|\kappa
^{f}(e^{\imath \lambda })|^{2}\right\vert =\lim_{\lambda \rightarrow
0}\lim_{T\rightarrow \infty }\left\vert |\kappa _{h}^{f}(e^{\imath \lambda
})|^{2}-|\kappa ^{f}(e^{\imath \lambda })|^{2}\right\vert $ and hence that $%
\nu _{2,T}=O(\delta _{T}^{2}h^{-2|d|})$.

The third and final term in \eqref{kkbar} is
\begin{equation}
|\kappa ^{f}(e^{\imath \lambda })|^{2}-|\kappa (e^{\imath \lambda
})|^{2}=|\kappa (e^{\imath \lambda })|^{2}\left( |1-e^{\imath \lambda
}|^{2(d_{T}^{f}-d)}-1\right) \,.  \label{kk3}
\end{equation}%
Substituting $|1-e^{\imath \lambda }|^{2(d_{T}^{f}-d)}=\exp
\{(d_{T}^{f}-d)\log |1-e^{\imath \lambda }|^{2}\}$ into \eqref{kk3} and
using the expansion $|1-e^{-\imath \lambda }|^{2}=2\sum_{j=1}^{\infty
}(-1)^{j-1}|\lambda |^{2j}/(2j)!$, which implies that $\log |1-e^{\imath
\lambda }|^{2}=2\log |\lambda |+\log (1+o(|\lambda |))$ as $\lambda
\rightarrow 0$, we can deduce that
\begin{equation*}
\left\vert |\kappa (e^{\imath \lambda })|^{2}(|1-e^{\imath \lambda
}|^{2(d_{T}^{f}-d)}-1)\right\vert \leq \{\sup_{[-\pi ,\pi ]}|\kappa
(e^{\imath \lambda })|^{2}\}\left\vert \exp \left\{ 2(d_{T}^{f}-d)\log
|\lambda |+o(|\lambda |)\right\} -1\right\vert
\end{equation*}%
as $\lambda \rightarrow 0$. Furthermore, by assumption $|d_{T}^{f}-d|\leq
\delta _{T}$ where $\delta _{T}\log T\rightarrow 0$ as $T\rightarrow \infty $%
, and since $|\exp (x)-1|=|x|\cdot |1+\half x+o(|x|)|$ for $x$ in a
neighbourhood of the origin, it follows that
\begin{equation*}
\left\vert |\kappa (e^{\imath \lambda })|^{2}(|1-e^{\imath \lambda
}|^{2(d_{T}^{f}-d)}-1)\right\vert \leq 2\{\sup_{[-\pi ,\pi ]}|\kappa
(e^{\imath \lambda })|^{2}\}\left\vert d_{T}^{f}-d\right\vert \left\vert
(\log 2\pi N/T)+o(N/T)\right\vert
\end{equation*}%
for all $\lambda \in \lbrack 2\pi /T,2\pi N/T]$ as $T\rightarrow \infty $.
We can therefore infer that \eqref{kk3} is $O(\delta _{T}\log T)$ or
smaller, uniformly in $\lambda $ for all $\lambda \in \lbrack 2\pi /T,2\pi
N/T]$. This completes the proof of Lemma \ref{kk}. $\hfill \qed$

\bigskip\noindent Returning to the proof of Theorem \ref{pfsbbias},
evaluating the expression
\begin{equation}
(\bar{c}_{0}-c_{0})+(\bar{c}_{1}-c_{1})|\lambda |^{2}=|\bar{\kappa}%
_{h}(e^{\imath \lambda })|^{2}-|\kappa (e^{\imath \lambda })|^{2}+o(|\lambda
|^{3})  \label{pfsbbias2}
\end{equation}%
at $\lambda =2\pi /T$ and $2\pi N/T$, and solving for $\bar{c}_{0}-c_{0}$
and $\bar{c}_{1}-c_{1}$, it follows as a consequence of Lemma \ref{kk} that $%
|\bar{c}_{0}-c_{0}|=O\left( M_{T}\right) +o(T^{-3})$ and $|\bar{c}%
_{1}-c_{1}|=O\left( T^{2}M_{T}/N^{2}\right) +o(N/T)$. Extracting the
dominant term completes the proof of Theorem \ref{pfsbbias}. $\hfill \qed$

\paragraph{Proof of Proposition \protect\ref{lrgdev}:}

Let $\widehat{d}_{T}$ denote the $\textit{LPR}(0)$ estimator. Then $\widehat{d}_{T}$
is the OLS coefficient of the regressor $-2\log \lambda _{j}$ in the
regression of $\log I_{T}(\lambda _{j})$ on $1$ and $-2\log \lambda _{j}$.
Substituting $a_{j}-2d\log (\lambda _{j})+\eta _{j}$ for $\log I_{T}(\lambda
_{j})$ in this regression leads to the expression
\begin{align}
\widehat{d}_{T}-d& =-\frac{\sum_{j=1}^{N}(\log \lambda _{j}-\overline{\log
\lambda })(\eta _{j}+a_{j})}{2\sum_{j=1}^{N}(\log \lambda _{j}-\overline{%
\log \lambda })^{2}}  \notag  \label{esterror} \\
& =-\half\sum_{j=1}^{N}r_{j}(\eta _{j}+a_{j})
\end{align}%
for the estimation error where $\eta _{j}$ and $a_{j}$ are defined in
expressions \eqref{lpr1} and \eqref{lpr2}, and $r_{j}=(\log \lambda _{j}-%
\overline{\log \lambda })/\sum_{j=1}^{N}(\log \lambda _{j}-\overline{\log
\lambda })^{2}$, $j=1,\ldots ,N$. See the discussion associated with %
\eqref{lpr1} and \eqref{lpr2} for clarification.

By Theorem 2 of \cite{moulines:soulier:1999} there exist sequences $e_{j}$
and $f_{j}$, $j=1,\ldots ,N$, such that $\eta _{j}=e_{j}+f_{j}$, where the $%
e_{j}$, $j=1,\ldots ,N$, are weakly dependent, centered Gumbel random
variables with variance $\pi ^{2}/6$ and covariance $cov\{e_{k},e_{j}\}=O(%
\log ^{2}(j)k^{-2|d|}j^{2(|d|-1)})$ for $1\leq k<j\leq N$, and $%
|f_{j}|=O(\log (1+j)/j)$ with probability one. Since $\max_{1\leq j\leq
N}|\log \lambda _{j}-\overline{\log \lambda }|=O(\log N)$ and $%
\sum_{j=1}^{N}(\log \lambda _{j}-\overline{\log \lambda })^{2}=O(N)$ it
follows that $\sum_{j=1}^{N}r_{j}f_{j}=O(\log ^{3}N/N)$ $a.s.$. Given that $%
\sum_{j=1}^{N}r_{j}=0$, it also follows from \eqref{lpr2b} that $%
\sum_{j=1}^{N}r_{j}a_{j}=O(N^{2}\log N/T^{2})$. We can therefore infer from %
\eqref{esterror} that
\begin{equation*}
\widehat{d}_{T}-d=-\half\sum_{j=1}^{N}r_{j}e_{j}+R_{N}
\end{equation*}%
where $|R_{N}|\log T=O(\nu ^{3}\log ^{4}T/T^{\nu })+O(\nu \log
^{2}T/T^{2(1-\nu )})=o(1)$ $a.s.$, $2/3<\nu <4/5$.

The desired result now follows because on application of a law of large
numbers for triangular arrays of weakly dependent random variables we find
that for all $\delta >0$
\begin{equation*}
\sum_{j=1}^{N}r_{j}e_{j}=o\left( (\nu \log T)^{5/2}(\log (\nu \log
T))^{(1+\delta )/2}T^{-\nu /2}\right) \quad a.s.\quad .
\end{equation*}%
More specifically, let $S_{n}=\sum_{j=1}^{n}r_{j}e_{j}$. Then by Doob's
inequality $E[(\max_{n\leq 2^{k}}|S_{n}|)^{2}]\leq 4E[|S_{2^{k}}|^{2}]$, and
using the bounds on the covariance of $e_{j}$ we have
\begin{equation*}
E[|S_{n}|^{2}]=\sum_{j=1}^{n}r_{j}^{2}E[e_{j}^{2}]+2\sum_{1\leq k<j\leq
n}r_{k}r_{j}cov\{e_{k},e_{j}\}=O(\log ^{4}n/n)\,.
\end{equation*}%
We can therefore conclude that for any $\delta >0$
\begin{equation*}
\sum_{k=1}^{\infty }\frac{2^{k}}{k^{5}(\log k)^{1+\delta }}E[(\max_{n\leq
2^{k}}|S_{n}|)^{2}]\leq \sum_{k=1}^{\infty }\frac{2^{k}}{k^{5}(\log
k)^{1+\delta }}\,O\left( \frac{k^{4}}{2^{k}}\right) <\infty \,,
\end{equation*}%
since $\sum_{k=1}^{\infty }1/k(\log k)^{1+\delta }<\infty $, which by the
Borel-Cantelli lemma implies $\max_{n\leq 2^{k}}|S_{n}|=o(k^{5/2}(\log
k)^{(1+\delta )/2}2^{-k/2})$ $a.s.$. Consequently $\sqrt{N}|S_{N}|=o((\log
N)^{5/2}(\log \log N)^{(1+\delta )/2})$ $a.s.$ since the function $(\log
n)^{5/2}(\log \log n)^{(1+\delta )/2}$ is slowly varying at infinity.

Now let $\widehat{d}_{T}$ denote the $\textit{LPR}(P)$ estimator with $P\geq
1$. The analytically-bias-adjusted LPR estimator is the OLS coefficient of
the regressor $-2\log \lambda _{j}$ in the regression of $\log I_{T}(\lambda
_{j})$ on $1$, $-2\log \lambda _{j}$, and $\lambda _{j}^{2p}$, $p=1,\ldots ,P
$. Applying the Frisch-Waugh-Lovell theorem and projecting out the
regressors $\lambda _{j}^{2p}$, $p=1,\ldots ,P$, as well as unity we can
express the estimation error $\widehat{d}_{T}-d$ exactly as in %
\eqref{esterror}, save that the $r_{j}$ are now defined in terms of $-2%
\widetilde{\log \lambda }_{j}$, say, the component of $-2\log \lambda _{j}$
orthogonal to $1$ and $\lambda _{j}^{2p}$, $p=1,\ldots ,P$. This projection
does not alter the overall magnitudes, so for the orthogonalized regressor
we have $\max_{1\leq j\leq N}|\widetilde{\log \lambda }_{j}|=O(\log N)$ and $%
\sum_{j=1}^{N}(\widetilde{\log \lambda }_{j})^{2}=O(N)$
\citep[][Lemma 2,
parts (j) \& (k)]{andrews:guggenberger:2003}. The proof that $|\widehat{d}%
_{T}-d|\log T=o(1)$ $a.s.$ now proceeds as previously with $r_{j}=\widetilde{%
\log \lambda }_{j}/\sum_{j=1}^{N}(\widetilde{\log \lambda }_{j})^{2}$, $%
j=1,\ldots ,N$.

For the $\textit{SPLW}(0)$ estimator the proposition follows directly from
\citet[][Lemma
5.8]{giraitis:robinson:2003}, which implies that the SPLW estimator
satisfies $P(|\widehat{d}_{T}-d|\log T>\epsilon )=o(N^{-p})$, where $%
p>1/\epsilon $ and $N$, the bandwidth, satisfies $T^{\epsilon
}<N<T^{1-\epsilon }$ for some $\epsilon >0$. For the $\textit{SPLW}(P\geq 1)$
estimator the proposition can be established in a manner similar to that
employed above for the $\textit{LPR}(P)$ estimators. Using Lemma 4 of \cite%
{andrew:sun:2004} we can express $\widehat{d}_{T}-d$, where $\widehat{d}_{T}$
now denotes the $\textit{SPLW}(P)$ estimator with $P\geq 1$, as a function
of the standardized score and from \citet[][Lemma 5]{andrew:sun:2004} we can
conclude that the standardized score is of an order that implies that $|%
\widehat{d}_{T}-d|\log T=o(1)$ $a.s.$; \textit{cf}.
\citet[][Theorem
4]{andrew:sun:2004}.\hfill \qed

%

\paragraph{Proof of Theorem \protect\ref{hpd_coverage}:}

The sampling distribution of $N^{1/2}(\widehat{d}_{T}-d)$ for the LPR and
SPLW estimators, under the Gaussianity assumption, admits a normal
approximation such that%
\begin{equation}
\sup_{x}\left\vert {\Pr }\left\{ \frac{N^{1/2}(\widehat{d}_{T}-d)}{%
\upsilon }<x\right\} -G(x)\right\vert =o\left( \frac{N^{5/2}}{T^{2}}\right) ,
\label{dEdge}
\end{equation}%
where $\upsilon =\omega \psi _{P}$ and $G(\cdot )$ denotes the standard
normal distribution function. Substituting \eqref{bt} into \eqref{dEdge}
gives the approximation
\begin{align}
{\Pr }\left\{ N^{1/2}\left( \widehat{d}_{T}-E[\widehat{d}%
_{T}]\right) <x\right\} & ={\Pr }\left\{ N^{1/2}\left( \widehat{d}%
_{T}-d\right) <x+N^{1/2}b_{T}\right\}  \notag \\
& =G\left( (x+N^{1/2}b_{T})/\upsilon \right) +o(N^{5/2}/T^{2})
\label{expansion}
\end{align}%
for the distribution of the finite sample deviation $\widehat{d}_{T}-E[%
\widehat{d}_{T}]$. Similarly,
\begin{equation}
\sup_{x}\left\vert {\Pr }^{\ast }\left\{ \frac{N^{1/2}(\widehat{d}%
_{T}^{\ast _{f}}-d_{T}^{f})}{\upsilon }<x\right\} -G(x)\right\vert =o\left(
\frac{N^{5/2}}{T^{2}}\right) \,,  \label{dEdge*}
\end{equation}%
and substituting \eqref{btbar} into \eqref{dEdge*} we obtain the
approximation 
\begin{align}
{\Pr }^{\ast }\left\{ N^{1/2}\left( \widehat{d}_{T}^{\ast
_{f}}-E^{\ast }[\widehat{d}_{T}^{\ast _{f}}]\right) <x\right\} & ={%
\Pr }^{\ast }\left\{ N^{1/2}\left( \widehat{d}_{T}^{\ast
_{f}}-d_{T}^{f}\right) <x+N^{1/2}b_{T}^{\ast }\right\}  \notag \\
& =G\left( \left( x+N^{1/2}b_{T}^{\ast }\right) /\upsilon \right)
+o(N^{5/2}/T^{2})  \label{bs_expansion}
\end{align}%
for the bootstrap deviation $\widehat{d}_{T}^{\ast _{f}}-E^{\ast }[\widehat{d%
}_{T}^{\ast _{f}}]$. Subtracting \eqref{expansion} from \eqref{bs_expansion}
we find that the difference between ${\Pr }^{\ast }\{N^{1/2}(%
\widehat{d}_{T}^{\ast _{f}}-E^{\ast }[\widehat{d}_{T}^{\ast _{f}}])<x\}$ and
${\Pr }\{N^{1/2}(\widehat{d}_{T}-E[\widehat{d}_{T}])<x\}$ is bounded
in absolute value by%
\begin{equation*}
\left\vert G\left( (x+N^{1/2}b_{T})/\upsilon \right)
-G((x+N^{1/2}b_{T}^{\ast })/\upsilon )\right\vert +o(N^{5/2}/T^{2})\,,
\end{equation*}%
and
\begin{equation*}
\sup_{x}\left\vert G((x+N^{1/2}b_{T})/\upsilon )-G((x+N^{1/2}b_{T}^{\ast
})/\upsilon )\right\vert \leq \frac{N^{1/2}}{\upsilon \sqrt{2\pi }}%
|b_{T}-b_{T}^{\ast }|
\end{equation*}%
by the first mean value theorem for integrals
\citep[][Theorem
7.30]{apostol:1960}.

Recognizing that $\overline{d}_{T,B}^{\ast _{f}}$ $=d_{T}^{f}+\widehat{b}%
_{T,B}^{\ast _{f}}$ and employing the expansion \eqref{dEdge*} once more we
also have the approximation
\begin{align}
{\Pr }^{\ast }\{N^{\half}(\widehat{d}_{T}^{\ast _{f}}-\overline{d}%
_{T,B}^{\ast _{f}})<x\}& ={\Pr }^{\ast }\left\{ N^{\half}(\widehat{d}%
_{T}^{\ast _{f}}-d_{T}^{f})<x+N^{\half}\widehat{b}_{T,B}^{\ast _{f}}\right\}
\notag \\
& =G\left( (x+N^{\half}\widehat{b}_{T,B}^{\ast _{f}})/\upsilon \right) +o(
N^{5/2}/T^{2})\,.  \label{approx 2}
\end{align}%
Subtracting the approximation in \eqref{approx 2} from that in %
\eqref{expansion} and using the triangle inequality, having added and
subtracted \eqref{bs_expansion}, now yields the result that
\begin{equation}
\sup_{x}\left\vert {\Pr }\left\{ N^{\half}(\widehat{d}_{T}-E[%
\widehat{d}_{T}])<x\right\} -{\Pr }^{\ast }\left\{ N^{\half}(%
\widehat{d}_{T}^{\ast _{f}}-\overline{d}_{T,B}^{\ast _{f}})<x\right\}
\right\vert \leq \frac{N^{\half}|b_{T}-b_{T}^{\ast }|}{\upsilon \sqrt{2\pi }}%
+r_{BN}  \label{approx 3}
\end{equation}%
where the remainder $r_{BN}=N^{\half}O( \sqrt{\log \log B/B}) +o(
N^{5/2}/T^{2})$. The first term on the right hand side in \eqref{approx 3}
follows from the inequality $\sup_{x}|G((x+N^{\half}b_{T})/\upsilon
)-G((x+N^{\half}b_{T}^{\ast })/\upsilon )|\leq N^{\half}|b_{T}-b_{T}^{\ast
}|/\upsilon \sqrt{2\pi }$, and similarly, the first term of the remainder
derives from the inequality $\sup_{x}|G((x+N^{\half}\widehat{b}_{T,B}^{\ast
_{f}})/\upsilon )-G((x+N^{\half}b_{T}^{\ast })/\upsilon )|\leq N^{\half}|%
\widehat{b}_{T,B}^{\ast _{f}}-b_{T}^{\ast }|/\upsilon \sqrt{2\pi }$, since $|%
\widehat{b}_{T,B}^{\ast _{f}}-b_{T}^{\ast }|=O(\sqrt{\log \log B/B})$ a.s..
Furthermore, if we set
\begin{equation*}
\overline{{\Pr }}^{\ast }\left\{ N^{\half}(\widehat{d}_{T}^{\ast
_{f}}-\overline{d}_{T,B}^{\ast _{f}})<x\right\} =B^{-1}\sum_{b=1}^{B}\mathbf{%
1}\left\{ N^{\half}(\widehat{d}_{T,b}^{\ast _{f}}-\overline{d}_{T,B}^{\ast
_{f}})\leq x\right\}
\end{equation*}%
then by the Dvoretsky--Kiefer--Wolfowitz inequality the probability of the
event
\begin{equation*}
\sup_{x}\left\vert \overline{{\Pr }}^{\ast }\left\{ N^{\half}(%
\widehat{d}_{T}^{\ast _{f}}-\overline{d}_{T,B}^{\ast _{f}})<x\right\} -%
{\Pr }^{\ast }\left\{ N^{\half}(\widehat{d}_{T}^{\ast _{f}}-%
\overline{d}_{T,B}^{\ast _{f}})<x\right\} \right\vert >\delta
\end{equation*}%
is bounded by $2\exp (-2B\delta ^{2})$. It therefore follows that for all $%
B\sim KT^{4/5+\beta }$, $\beta >0$, the $(1-\alpha _{U}-\alpha _{L})100\%$
significance level HPD intervals are consistent with respect to the
Kolmogorov-Smirnov metric. $\hfill \qed$

\section{Stochastic Stopping Rules\label{ssr}}

Two criteria are used to determine if any meaningful gain in accuracy will
be achieved by adding a further iteration to the iterative procedure of
Section \ref{recursive}. The first, $|\widetilde{d}_{T}^{(k+1)}-\widetilde{d}%
_{T}^{(k)}|>\tau _{1}^{(k)}$, is based on Cauchy's convergence criterion.
Given the stochastic nature of the bias correction mechanism we can think of
this as a statistical decision rule in which $\tau _{1}^{(k)}$ governs the
probability of moving from the $k^{th}$ to the $(k+1)^{th}$ iteration. Now $%
\widetilde{d}_{T}^{(k+1)}-\widetilde{d}_{T}^{(k)}=-\widetilde{b}_{T,B}^{\ast
_{f}(k)}=-\frac{1}{B}\sum_{b=1}^{B}\left( \tilde{d}_{T,b}^{\ast _{f}(k)}-%
\widetilde{d}_{T}^{(k)}\right) $, where $\tilde{d}_{T,b}^{\ast _{f}(k)}$\
denotes the estimator produced from the $b^{th}$ PFSB draw with $\widetilde{d%
}_{T}^{(k)}$ used as the pre-filtering value; and since $\widehat{d}_{T}$ is
a $\sqrt{N}$--CAN estimator, given the data and the current and previous
bootstrap iterations, $N^{\half}(\tilde{d}_{T,b}^{\ast _{f}(k)}-\widetilde{d}%
_{T}^{(k)})\overset{\mathcal{D}}{\rightarrow }\mathcal{N}(0,\upsilon ^{2})$.
The conditional (asymptotic) variance of $B^{-1}\sum_{b=1}^{B}\left( \tilde{d%
}_{T,b}^{\ast _{f}(k)}-\widetilde{d}_{T}^{(k)}\right) $ is therefore $%
\upsilon ^{2}/NB$, and using the rule that the overall variance equals the
variance of the conditional mean (in this case $Var[\widetilde{d}_{T}^{(k)}]$%
) plus the expectation of the conditional variance (in this case the
constant $\upsilon ^{2}/NB$) we can infer that the (asymptotic) variance of
the difference between successive bias-adjusted estimators is given by $Var%
\left[ \widetilde{d}_{T}^{(k+1)}-\widetilde{d}_{T}^{(k)}\right] =Var\left[
\widetilde{d}_{T}^{(k)}\right] +\frac{\upsilon ^{2}}{NB}\,.$ Furthermore,
from the recurrence formula $\widetilde{d}_{T}^{(k)}=\widetilde{d}%
_{T}^{(k-1)}-\widetilde{b}_{T,B}^{\ast _{f}(k-1)}=\widetilde{d}_{T}^{(k-1)}-%
\frac{1}{B}\sum_{b=1}^{B}\left( \tilde{d}_{T,b}^{\ast _{f}(k-1)}-\widetilde{d%
}_{T}^{(k-1)}\right) $, it follows by a similar logic that $Var\left[
\widetilde{d}_{T}^{(k)}\right] =2\cdot Var\left[ \widetilde{d}_{T}^{(k-1)}%
\right] +\frac{\upsilon ^{2}}{NB},$ where $Var[\widetilde{d}%
_{T}^{(1)}]=2\cdot Var[\widetilde{d}_{T}^{(0)}]+\upsilon
^{2}/NB=(2B+1)\upsilon ^{2}/NB$. Moreover, at each iteration the
bias-adjusted estimate is constructed as a linear combination of
asymptotically normal random variables and is itself therefore
asymptotically normal. This indicates that $\tau _{1}^{(k)}$ can be
evaluated from percentile points of the normal approximation.

Similarly, the second convergence criterion, $\left\vert \widetilde{d}%
_{T}^{(0)}-\widetilde{d}_{T}^{(k)}-\widetilde{b}_{T,B}^{\ast
_{f}(k)}\right\vert >\tau _{2}^{(k)}$, is perhaps best thought of as the
decision rule that examines the difference between the current accumulated
bias correction, $\widetilde{d}_{T}^{(0)}-\widetilde{d}_{T}^{(k)}$, and the
current bootstrap estimate of the bias, $\widetilde{b}_{T,B}^{\ast _{f}(k)}$. 
From the expression $\widetilde{d}_{T}^{(0)}-\widetilde{d}_{T}^{(k)}-%
\widetilde{b}_{T,B}^{\ast _{f}(k)}=\widetilde{d}_{T}^{(0)}-\left( \frac{1}{B}%
\sum\limits_{b=1}^{B}\tilde{d}_{T,b}^{\ast _{f}(k)}\right) ,$ it follows
that the (asymptotic) variance, $Var\left[ \widetilde{d}_{T}^{(0)}-%
\widetilde{d}_{T}^{(k)}-\widetilde{b}_{T,B}^{\ast _{f}(k)}\right] =\frac{%
\upsilon ^{2}}{N}\left( 1+2^{k-1}\left[ 1+\frac{1}{B}\right] \right) \,,$
and the tolerance level $\tau _{2}^{(k)}$ can once again be set using
percentile points from the asymptotic normal approximation.

The interpretation of the convergence criteria as statistical decision rules
in which the tolerance levels govern the probability of going from the
current to the next iteration suggests that $\tau _{1}^{(k)}$ and $\tau
_{2}^{(k)}$ be set by reference to conventional critical values used in
statistical hypothesis tests. When $k$ is very small we might conjecture
that $\widetilde{d}_{T}^{(k)}$ still contains some bias and we may wish to
iterate further unless there is strong evidence that so doing will produce
very little change. On the other hand, when $k$ is large the initial
estimate $\widetilde{d}_{T}^{(0)}$ has already undergone several adjustments
to produce $\widetilde{d}_{T}^{(k)}$ and we may prefer to terminate
iteration unless there is strong evidence that further iteration will
produce additional, substantial correction. We can therefore calibrate $\tau
_{1}^{(k)}$ and $\tau _{2}^{(k)}$ using quantile points of the normal
distribution $z_{(1-p_{k}/2)}$ (where $G(z_{(1-p)})=1-p$) and $p_{k}$, the
probability of going from the $k^{th}$ to the $(k+1)^{th}$ iteration, is
assigned to be large when $k$ is small and vice versa. In the simulation
experiments we set $p_{0}=0.95$, $p_{1}=0.9$, and $p_{k}=(0.1)2^{(1-k)}$ for
$k=2,3,\ldots $\thinspace for $\textit{LPR}(0)$ and $\textit{SPLW}(0)$; and $p_{0}=0.9$, $%
p_{k}=(0.1)2^{-k}$ for $k=1,2,3,\ldots \,$ for $\textit{LPR}(P)$ and $%
\textit{SPLW}(P)$, $P\geq 1$.

\section{Tables\label{tbls}}


\begin{landscape}%

\begin{table}[tbp] \centering%
\thinmuskip=2mu\medmuskip=0mu\thickmuskip=2mu
\caption{Bias and mean square error (MSE) for all LPR-based estimators; $T=100$, using the parametric version of the bootstrap.\\
(Unadjusted and analytically-bias-adjusted $\textit{LPR}(P)$, $P=0,1,2$; plus bootstrap-bias-adjusted\\
$\textit{LPR}(P)\textit{-BBA}(K)$; $K=1,\ldots,3-P$, and their iteratively-adjusted (\textit{SSR}) variants.)\\
The lowest bias (in absolute value) and MSE  for each design are highlighted.\\
Analogous figures for the ML estimator are reported in italics.}\label%
{table-lpr-100}

\begin{tabular}{ccccccccccccccc}
&  & \multicolumn{3}{c}{$\textit{LPR}(P)$} & \multicolumn{4}{c}{$\textit{LPR}%
(0)\textit{-BBA}(K)$} & \multicolumn{3}{c}{$\textit{LPR}(1)\textit{-BBA}(K)$}
& \multicolumn{2}{r}{$\textit{LPR}(2)\textit{-BBA}(K)$} & $(Gaussian)$ \\%
[0.5ex]
&  & $P=0$ & $P=1$ & $P=2$ & \multicolumn{1}{|c}{$K=1$} & $K=2$ & $K=3$ & $%
SSR$ & \multicolumn{1}{|c}{$K=1$} & $K=2$ & $SSR$ & \multicolumn{1}{|c}{$K=1$%
} & $SSR$ & \multicolumn{1}{|c}{$ML$} \\ \hline
&  &  &  &  &  &  &  &  &  &  &  &  &  &  \\[-0.5ex]
$d$ & $\phi $ & \multicolumn{13}{c}{Bias} \\[1ex]
0 & 0.3 & 0.1391 & 0.0329 & 0.0145 & 0.1210 & 0.0909 & 0.0320 & 0.1142 &
0.0134 & -0.0241 & \cellcolor{hcolor}{-0.0006} & -0.0156 & -0.0366 & \textit{%
-0.0933} \\
& 0.6 & 0.3873 & 0.2062 & 0.0929 & 0.3387 & 0.2589 & 0.1169 & 0.3215 & 0.1615
& 0.0873 & 0.1477 & 0.0445 & \cellcolor{hcolor}{0.0338} & \textit{-0.0321}
\\
& 0.9 & 0.8141 & 0.7393 & 0.6352 & 0.7935 & 0.7592 & 0.6628 & 0.7849 & 0.7014
& 0.6189 & 0.6836 & 0.5922 & \cellcolor{hcolor}{0.5706} & \textit{0.1222} \\
0.2 & 0.3 & 0.1312 & 0.0376 & 0.0124 & 0.1123 & 0.0814 & 0.0240 & 0.1015 &
0.0174 & -0.0155 & \cellcolor{hcolor}{0.0082} & -0.0165 & -0.0280 & \textit{%
-0.1055} \\
& 0.6 & 0.3790 & 0.2037 & 0.1237 & 0.3263 & 0.2402 & 0.0811 & 0.2990 & 0.1514
& 0.0669 & 0.1372 & 0.0744 & \cellcolor{hcolor}{0.0556} & \textit{-0.0652}
\\
& 0.9 & 0.7955 & 0.7237 & 0.6414 & 0.8166 & 0.8164 & 0.6524 & 0.8024 & 0.7232
& \cellcolor{hcolor}{0.6161} & 0.7042 & 0.6358 & 0.6168 & \textit{0.0330} \\
0.4 & 0.3 & 0.1335 & 0.0420 & 0.0162 & 0.1066 & 0.0636 & -0.0181 & 0.0920 &
0.0203 & -0.0183 & \cellcolor{hcolor}{0.0126} & -0.0158 & -0.0360 & \textit{%
-0.1103} \\
& 0.6 & 0.3720 & 0.2130 & 0.1142 & 0.3214 & 0.2360 & 0.0705 & 0.3031 & 0.1623
& 0.0716 & 0.1435 & 0.0709 & \cellcolor{hcolor}{0.0619} & \textit{-0.0937}
\\
& 0.9 & 0.7144 & 0.6768 & 0.6165 & 0.7939 & 0.7504 & %
\cellcolor{hcolor}{0.4633} & 0.7607 & 0.7269 & 0.4985 & 0.6780 & 0.6562 &
0.6312 & \textit{-0.0161} \\
&  &  &  &  &  &  &  &  &  &  &  &  &  &  \\[-0.5ex]
$d$ & $\phi $ & \multicolumn{13}{c}{MSE} \\[1ex]
0 & 0.3 & \cellcolor{hcolor}{0.0422} & 0.0804 & 0.1563 & 0.0591 & 0.1270 &
0.3720 & 0.0822 & 0.1349 & 0.2869 & 0.1858 & 0.2743 & 0.3556 & \textit{0.0549%
} \\
& 0.6 & 0.1754 & \cellcolor{hcolor}{0.1107} & 0.1479 & 0.1649 & 0.2019 &
0.4207 & 0.2027 & 0.1463 & 0.2855 & 0.1897 & 0.2413 & 0.2995 & \textit{0.0367%
} \\
& 0.9 & 0.6866 & 0.6179 & \cellcolor{hcolor}{0.5495} & 0.6781 & 0.7006 &
0.7788 & 0.6870 & 0.6249 & 0.6659 & 0.6556 & 0.6202 & 0.6842 & \textit{0.0474%
} \\
0.2 & 0.3 & \cellcolor{hcolor}{0.0422} & 0.0800 & 0.1547 & 0.0617 & 0.1383 &
0.3968 & 0.1010 & 0.1346 & 0.2767 & 0.1704 & 0.2488 & 0.3110 & \textit{0.0591%
} \\
& 0.6 & 0.1691 & \cellcolor{hcolor}{0.1145} & 0.1600 & 0.1581 & 0.1986 &
0.4346 & 0.2155 & 0.1480 & 0.2873 & 0.1860 & 0.2445 & 0.3190 & \textit{0.0340%
} \\
& 0.9 & 0.6592 & 0.5977 & \cellcolor{hcolor}{0.5565} & 0.7405 & 0.9021 &
1.0148 & 0.7863 & 0.6953 & 0.7698 & 0.7232 & 0.7146 & 0.7622 & \textit{0.0185%
} \\
0.4 & 0.3 & \cellcolor{hcolor}{0.0417} & 0.0718 & 0.1654 & 0.0558 & 0.1171 &
0.3411 & 0.0974 & 0.1167 & 0.2516 & 0.1551 & 0.2746 & 0.3672 & \textit{0.0488%
} \\
& 0.6 & 0.1608 & \cellcolor{hcolor}{0.1175} & 0.1566 & 0.1523 & 0.1918 &
0.4158 & 0.1865 & 0.1592 & 0.3146 & 0.2139 & 0.2446 & 0.3009 & \textit{0.0298%
} \\
& 0.9 & 0.5387 & 0.5401 & \cellcolor{hcolor}{0.5312} & 0.7325 & 1.0400 &
1.1629 & 0.8361 & 0.7686 & 0.8741 & 0.8838 & 0.8340 & 0.9044 & \textit{0.0084%
} \\
&  &  &  &  &  &  &  &  &  &  &  &  &  &
\end{tabular}
\end{table}%

\begin{table}[tbp] \centering%

\thinmuskip=2mu\medmuskip=0mu\thickmuskip=2mu
\caption{Bias and mean square error (MSE) for all LPR-based estimators; $T=500$, using the parametric version of the bootstrap.\\
(Unadjusted and analytically-bias-adjusted $\textit{LPR}(P)$, $P=0,1,2$; plus bootstrap-bias-adjusted\\
$\textit{LPR}(P)\textit{-BBA}(K)$; $K=1,\ldots,3-P$, and their iteratively-adjusted (\textit{SSR}) variants.)\\
The lowest bias (in absolute value) and MSE  for each design are highlighted.\\
Analogous figures for the ML estimator are reported in italics.}\label%
{table-lpr-500}

\begin{tabular}{ccccccccccccccc}
&  & \multicolumn{3}{c}{$\textit{LPR}(P)$} & \multicolumn{4}{c}{$\textit{LPR}%
(0)\textit{-BBA}(K)$} & \multicolumn{3}{c}{$\textit{LPR}(1)\textit{-BBA}(K)$}
& \multicolumn{2}{r}{$\textit{LPR}(2)\textit{-BBA}(K)$} & $(Gaussian)$ \\%
[0.5ex]
&  & $P=0$ & $P=1$ & $P=2$ & \multicolumn{1}{|c}{$K=1$} & $K=2$ & $K=3$ & $%
SSR$ & \multicolumn{1}{|c}{$K=1$} & $K=2$ & $SSR$ & \multicolumn{1}{|c}{$K=1$%
} & $SSR$ & \multicolumn{1}{|c}{$ML$} \\ \hline
&  &  &  &  &  &  &  &  &  &  &  &  &  &  \\[-0.5ex]
$d$ & $\phi $ & \multicolumn{13}{c}{Bias} \\[1ex]
0 & 0.3 & 0.0596 & 0.0081 & -0.0072 & 0.0317 & -0.0075 & -0.0709 & 0.0280 & %
\cellcolor{hcolor}{0.0013} & -0.0099 & -0.0023 & -0.0159 & -0.0239 & \textit{%
-0.0240} \\
& 0.6 & 0.2199 & 0.0705 & 0.0158 & 0.1558 & 0.0556 & -0.1176 & 0.1247 &
0.0332 & -0.0245 & 0.0239 & \cellcolor{hcolor}{-0.0103} & -0.0222 & \textit{%
-0.0276} \\
& 0.9 & 0.6722 & 0.4894 & 0.3619 & 0.5914 & 0.4610 & %
\cellcolor{hcolor}{0.2321} & 0.5441 & 0.4070 & 0.2782 & 0.3715 & 0.2831 &
0.2580 & \textit{0.0716} \\
0.2 & 0.3 & 0.0571 & 0.0083 & 0.0047 & 0.0310 & -0.0049 & -0.0628 & 0.0287 & %
\cellcolor{hcolor}{0.0003} & -0.0133 & -0.0057 & -0.0054 & -0.0115 & \textit{%
-0.0249} \\
& 0.6 & 0.2177 & 0.0702 & 0.0179 & 0.1532 & 0.0535 & -0.1195 & 0.1335 &
0.0299 & -0.0312 & 0.0224 & \cellcolor{hcolor}{-0.0091} & -0.0175 & \textit{%
-0.0321} \\
& 0.9 & 0.6670 & 0.4895 & 0.3667 & 0.5956 & 0.4752 & 0.2620 & 0.5360 & 0.3954
& 0.2486 & 0.3481 & 0.2814 & \cellcolor{hcolor}{0.2480} & \textit{0.0087} \\
0.4 & 0.3 & 0.0639 & 0.0206 & 0.0100 & 0.0353 & \cellcolor{hcolor}{-0.0036}
& -0.0654 & 0.0323 & 0.0091 & -0.0064 & 0.0054 & -0.0045 & -0.0101 & \textit{%
-0.0276} \\
& 0.6 & 0.2206 & 0.0761 & 0.0320 & 0.1480 & 0.0370 & -0.1557 & 0.1190 &
0.0298 & -0.0368 & 0.0190 & \cellcolor{hcolor}{0.0009} & -0.0048 & \textit{%
-0.0445} \\
& 0.9 & 0.6472 & 0.4872 & 0.3689 & 0.6503 & 0.6581 & 0.6030 & 0.6262 & 0.4190
& 0.3090 & 0.3904 & 0.2856 & \cellcolor{hcolor}{0.2444} & \textit{0.0023} \\
&  &  &  &  &  &  &  &  &  &  &  &  &  &  \\[-0.5ex]
$d$ & $\phi $ & \multicolumn{13}{c}{MSE} \\[1ex]
0 & 0.3 & \cellcolor{hcolor}{0.0101} & 0.0163 & 0.0294 & 0.0131 & 0.0282 &
0.0859 & 0.0210 & 0.0237 & 0.0434 & 0.0365 & 0.0404 & 0.0639 & \textit{0.0105%
} \\
& 0.6 & 0.0554 & \cellcolor{hcolor}{0.0224} & 0.0337 & 0.0390 & 0.0444 &
0.1470 & 0.0738 & 0.0309 & 0.0664 & 0.0550 & 0.0532 & 0.0907 & \textit{0.0159%
} \\
& 0.9 & 0.4581 & 0.2564 & 0.1591 & 0.3613 & 0.2424 & 0.1510 & 0.3678 & 0.1961
& 0.1517 & 0.2391 & \cellcolor{hcolor}{0.1326} & 0.1832 & \textit{0.0277} \\
0.2 & 0.3 & \cellcolor{hcolor}{0.0099} & 0.0157 & 0.0286 & 0.0134 & 0.0292 &
0.0849 & 0.0156 & 0.0223 & 0.0411 & 0.0441 & 0.0382 & 0.0610 & \textit{0.0105%
} \\
& 0.6 & 0.0541 & \cellcolor{hcolor}{0.0214} & 0.0283 & 0.0366 & 0.0391 &
0.1308 & 0.0573 & 0.0291 & 0.0641 & 0.0476 & 0.0432 & 0.0766 & \textit{0.0143%
} \\
& 0.9 & 0.4514 & 0.2571 & 0.1645 & 0.3695 & 0.2665 & 0.1974 & 0.4140 & 0.1904
& 0.1477 & 0.2609 & \cellcolor{hcolor}{0.1329} & 0.1957 & \textit{0.0055} \\
0.4 & 0.3 & \cellcolor{hcolor}{0.0110} & 0.0168 & 0.0299 & 0.0139 & 0.0288 &
0.0832 & 0.0183 & 0.0226 & 0.0394 & 0.0401 & 0.0383 & 0.0686 & \textit{0.0071%
} \\
& 0.6 & 0.0551 & \cellcolor{hcolor}{0.0235} & 0.0290 & 0.0342 & 0.0349 &
0.1329 & 0.0696 & 0.0292 & 0.0606 & 0.0538 & 0.0406 & 0.0618 & \textit{0.0110%
} \\
& 0.9 & 0.4259 & 0.2538 & 0.1672 & 0.4469 & 0.5214 & 0.6908 & 0.5318 & 0.2129
& 0.1944 & 0.2463 & \cellcolor{hcolor}{0.1441} & 0.2341 & \textit{0.0034} \\
&  &  &  &  &  &  &  &  &  &  &  &  &  &
\end{tabular}
\end{table}%

\begin{table}[tbp] \centering%
\thinmuskip=2mu\medmuskip=0mu\thickmuskip=2mu
\caption{Bias and mean square error (MSE) for all SPLW-based estimators; $T=100$, using the parametric version of the bootstrap.\\
(Unadjusted and analytically-bias-adjusted $\textit{SPLW}(P)$, $P=0,1,2$; plus bootstrap-bias-adjusted\\
$\textit{SPLW}(P)\textit{-BBA}(K)$; $K=1,\ldots,3-P$, and their iteratively-adjusted (\textit{SSR}) variants.)\\
The lowest bias (in absolute value) and MSE  for each design are highlighted.\\
Analogous figures for the ML estimator are reported in italics.}\label%
{table-lpw-100}

\begin{tabular}{ccccccccccccccc}
&  & \multicolumn{3}{c}{$\textit{SPLW}(P)$} & \multicolumn{4}{c}{$\textit{%
SPLW}(0)\textit{-BBA}(K)$} & \multicolumn{3}{c}{$\textit{SPLW}(1)\textit{-BBA%
}(K)$} & \multicolumn{2}{r}{$\textit{SPLW}(2)\textit{-BBA}(K)$} & $%
(Gaussian) $ \\[0.5ex]
&  & $P=0$ & $P=1$ & $P=2$ & \multicolumn{1}{|c}{$K=1$} & $K=2$ & $K=3$ & $%
SSR$ & \multicolumn{1}{|c}{$K=1$} & $K=2$ & $SSR$ & \multicolumn{1}{|c}{$K=1$%
} & $SSR$ & \multicolumn{1}{|c}{$ML$} \\ \hline
&  &  &  &  &  &  &  &  &  &  &  &  &  &  \\[-0.5ex]
$d$ & $\phi $ & \multicolumn{13}{c}{Bias} \\[1ex]
0 & 0.3 & 0.1300 & -0.0162 & -0.0648 & 0.1136 & 0.0901 & 0.0479 & 0.1124 & %
\cellcolor{hcolor}{-0.0010} & 0.0099 & -0.0037 & -0.0577 & -0.0760 & \textit{%
-0.0933} \\
& 0.6 & 0.3985 & 0.1583 & 0.0604 & 0.3695 & 0.3255 & 0.2500 & 0.3685 & 0.1431
& 0.1126 & 0.1394 & 0.0611 & \cellcolor{hcolor}{0.0496} & \textit{-0.0321}
\\
& 0.9 & 0.8242 & 0.7197 & 0.6122 & 0.8197 & 0.8120 & 0.7893 & 0.8172 & 0.7248
& 0.7190 & 0.7204 & 0.6123 & \cellcolor{hcolor}{0.5951} & \textit{0.1222} \\
0.2 & 0.3 & 0.1191 & \cellcolor{hcolor}{0.0011} & -0.0657 & 0.1063 & 0.0875
& 0.0545 & 0.1062 & 0.0192 & 0.0342 & 0.0162 & -0.0492 & -0.0537 & \textit{%
-0.1055} \\
& 0.6 & 0.3973 & 0.1602 & 0.0526 & 0.3607 & 0.3037 & 0.2033 & 0.3585 & 0.1424
& 0.1125 & 0.1386 & 0.0417 & \cellcolor{hcolor}{0.0325} & \textit{-0.0652}
\\
& 0.9 & 0.7944 & 0.6898 & 0.6012 & 0.8244 & 0.8615 & 0.8259 & 0.8301 & 0.7011
& 0.6645 & 0.6861 & 0.6181 & \cellcolor{hcolor}{0.5999} & \textit{0.0330} \\
0.4 & 0.3 & 0.1217 & -0.0090 & -0.0200 & 0.1044 & 0.0783 & 0.0329 & 0.1045 &
-0.0029 & \cellcolor{hcolor}{-0.0019} & -0.0062 & -0.0044 & -0.0092 &
\textit{-0.1103} \\
& 0.6 & 0.3828 & 0.1766 & 0.0425 & 0.3539 & 0.3072 & 0.2207 & 0.3495 & 0.1564
& 0.1208 & 0.1417 & 0.0324 & \cellcolor{hcolor}{0.0217} & \textit{-0.0937}
\\
& 0.9 & 0.7409 & 0.6762 & \cellcolor{hcolor}{0.5966} & 0.836 & 0.8916 &
0.7652 & 0.8669 & 0.7456 & 0.6859 & 0.7278 & 0.6619 & 0.6467 & \textit{%
-0.0161} \\
&  &  &  &  &  &  &  &  &  &  &  &  &  &  \\[-0.5ex]
$d$ & $\phi $ & \multicolumn{13}{c}{MSE} \\[1ex]
0 & 0.3 & \cellcolor{hcolor}{0.0329} & 0.0527 & 0.1133 & 0.0361 & 0.0577 &
0.1442 & 0.0387 & 0.0774 & 0.1508 & 0.0888 & 0.1758 & 0.2432 & \textit{0.0549%
} \\
& 0.6 & 0.1773 & \cellcolor{hcolor}{0.0803} & 0.1126 & 0.1611 & 0.1555 &
0.2015 & 0.1614 & 0.1039 & 0.1795 & 0.1170 & 0.1628 & 0.2089 & \textit{0.0367%
} \\
& 0.9 & 0.6975 & 0.5695 & \cellcolor{hcolor}{0.4973} & 0.7032 & 0.7287 &
0.8036 & 0.7019 & 0.6022 & 0.6551 & 0.6066 & 0.5650 & 0.5935 & \textit{0.0474%
} \\
0.2 & 0.3 & \cellcolor{hcolor}{0.0308} & 0.0492 & 0.1190 & 0.0348 & 0.0562 &
0.1374 & 0.0354 & 0.0745 & 0.1430 & 0.0835 & 0.1660 & 0.1994 & \textit{0.0591%
} \\
& 0.6 & 0.1770 & \cellcolor{hcolor}{0.0807} & 0.1081 & 0.1576 & 0.1501 &
0.2038 & 0.1583 & 0.0986 & 0.1626 & 0.1114 & 0.1540 & 0.1861 & \textit{0.0340%
} \\
& 0.9 & 0.6500 & 0.5346 & \cellcolor{hcolor}{0.4757} & 0.7247 & 0.8792 &
1.0641 & 0.7632 & 0.6079 & 0.6862 & 0.6358 & 0.5873 & 0.6211 & \textit{0.0185%
} \\
0.4 & 0.3 & \cellcolor{hcolor}{0.0315} & 0.0576 & 0.1134 & 0.0337 & 0.0529 &
0.1312 & 0.0337 & 0.0819 & 0.1491 & 0.0905 & 0.1562 & 0.1834 & \textit{0.0488%
} \\
& 0.6 & 0.1643 & \cellcolor{hcolor}{0.0869} & 0.1195 & 0.1533 & 0.1585 &
0.2347 & 0.1571 & 0.1074 & 0.1804 & 0.1531 & 0.1763 & 0.2100 & \textit{0.0298%
} \\
& 0.9 & 0.5672 & 0.5101 & \cellcolor{hcolor}{0.4593} & 0.7580 & 1.1781 &
1.4480 & 0.8734 & 0.6878 & 0.8896 & 0.7243 & 0.6727 & 0.7178 & \textit{0.0084%
} \\
&  &  &  &  &  &  &  &  &  &  &  &  &  &
\end{tabular}%
\end{table}%

\begin{table}[tbp] \centering%
\thinmuskip=2mu\medmuskip=0mu\thickmuskip=2mu
\caption{Bias and mean square error (MSE) for all SPLW-based estimators; $T=500$, using the parametric version of the bootstrap.\\
(Unadjusted and analytically-bias-adjusted $\textit{SPLW}(P)$, $P=0,1,2$; plus bootstrap-bias-adjusted\\
$\textit{SPLW}(P)\textit{-BBA}(K)$; $K=1,\ldots,3-P$, and their iteratively-adjusted (\textit{SSR}) variants.)\\
The lowest bias (in absolute value) and MSE  for each design are highlighted.\\
Analogous figures for the ML estimator are reported in italics.}\label%
{table-lpw-500}

\begin{tabular}{ccccccccccccccc}
&  & \multicolumn{3}{c}{$\textit{SPLW}(P)$} & \multicolumn{4}{c}{$\textit{%
SPLW}(0)\textit{-BBA}(K)$} & \multicolumn{3}{c}{$\textit{SPLW}(1)\textit{-BBA%
}(K)$} & \multicolumn{2}{r}{$\textit{SPLW}(2)\textit{-BBA}(K)$} & $%
(Gaussian) $ \\[0.5ex]
&  & $P=0$ & $P=1$ & $P=2$ & \multicolumn{1}{|c}{$K=1$} & $K=2$ & $K=3$ & $%
SSR$ & \multicolumn{1}{|c}{$K=1$} & $K=2$ & $SSR$ & \multicolumn{1}{|c}{$K=1$%
} & $SSR$ & \multicolumn{1}{|c}{$ML$} \\ \hline
&  &  &  &  &  &  &  &  &  &  &  &  &  &  \\[-0.5ex]
$d$ & $\phi $ & \multicolumn{13}{c}{Bias} \\[1ex]
0 & 0.3 & 0.0561 & -0.0137 & -0.0097 & 0.0289 & -0.0081 & -0.0638 & 0.0288 &
-0.0096 & -0.0064 & -0.0109 & 0.0028 & \cellcolor{hcolor}{0.0008} & \textit{%
-0.0240} \\
& 0.6 & 0.2305 & 0.0538 & 0.0040 & 0.1746 & 0.0892 & -0.0552 & 0.1639 &
0.0269 & -0.0130 & 0.0241 & \cellcolor{hcolor}{-0.0017} & -0.0056 & \textit{%
-0.0276} \\
& 0.9 & 0.7220 & 0.5220 & 0.3798 & 0.6731 & 0.6003 & 0.4808 & 0.6694 & 0.4683
& 0.3878 & 0.4628 & 0.3270 & \cellcolor{hcolor}{0.3158} & \textit{0.0716} \\
0.2 & 0.3 & 0.0566 & -0.0077 & -0.0100 & 0.0323 & -0.0002 & -0.0481 & 0.0323
& -0.0006 & 0.0066 & -0.0025 & \cellcolor{hcolor}{0.0000} & -0.0042 &
\textit{-0.0249} \\
& 0.6 & 0.2291 & 0.0494 & 0.0046 & 0.1731 & 0.0892 & -0.0539 & 0.1618 &
0.0204 & -0.0226 & 0.0190 & \cellcolor{hcolor}{-0.0011} & -0.0041 & \textit{%
-0.0321} \\
& 0.9 & 0.7177 & 0.5220 & 0.3838 & 0.6829 & 0.6278 & 0.5347 & 0.6799 & 0.4648
& 0.3797 & 0.4599 & 0.3242 & \cellcolor{hcolor}{0.3132} & \textit{0.0087} \\
0.4 & 0.3 & 0.0603 & 0.0056 & -0.0086 & 0.0338 & \cellcolor{hcolor}{-0.0013}
& -0.0525 & 0.0336 & 0.0093 & 0.0125 & 0.0077 & -0.0015 & -0.0049 & \textit{%
-0.0276} \\
& 0.6 & 0.2265 & 0.0572 & \cellcolor{hcolor}{0.0046} & 0.1623 & 0.0662 &
-0.0977 & 0.1440 & 0.0242 & -0.0218 & 0.0245 & -0.0104 & -0.0090 & \textit{%
-0.0445} \\
& 0.9 & 0.6982 & 0.5093 & 0.3819 & 0.7399 & 0.8157 & 0.8543 & 0.7850 & 0.4689
& 0.4047 & 0.4583 & 0.3267 & \cellcolor{hcolor}{0.3180} & \textit{0.0023} \\
&  &  &  &  &  &  &  &  &  &  &  &  &  &  \\[-0.5ex]
$d$ & $\phi $ & \multicolumn{13}{c}{MSE} \\[1ex]
0 & 0.3 & \cellcolor{hcolor}{0.0075} & 0.0118 & 0.0196 & 0.0078 & 0.0141 &
0.0382 & 0.0080 & 0.0159 & 0.0248 & 0.0158 & 0.0233 & 0.0233 & \textit{0.0105%
} \\
& 0.6 & 0.0577 & \cellcolor{hcolor}{0.0135} & 0.0193 & 0.0374 & 0.0246 &
0.0552 & 0.0432 & 0.0175 & 0.0335 & 0.0242 & 0.0283 & 0.0347 & \textit{0.0159%
} \\
& 0.9 & 0.5264 & 0.2867 & 0.1663 & 0.4603 & 0.3740 & 0.2657 & 0.4574 & 0.2406
& 0.1920 & 0.2454 & \cellcolor{hcolor}{0.1422} & 0.1589 & \textit{0.0277} \\
0.2 & 0.3 & \cellcolor{hcolor}{0.0075} & 0.0117 & 0.0217 & 0.0076 & 0.0129 &
0.0325 & 0.0076 & 0.0151 & 0.0224 & 0.0150 & 0.0256 & 0.0286 & \textit{0.0105%
} \\
& 0.6 & 0.0572 & \cellcolor{hcolor}{0.0140} & 0.0204 & 0.0368 & 0.0246 &
0.0555 & 0.0429 & 0.0183 & 0.0347 & 0.0217 & 0.0282 & 0.0314 & \textit{0.0143%
} \\
& 0.9 & 0.5205 & 0.2866 & 0.1701 & 0.4768 & 0.4173 & 0.3460 & 0.4756 & 0.2373
& 0.1854 & 0.2384 & \cellcolor{hcolor}{0.1411} & 0.1575 & \textit{0.0055} \\
0.4 & 0.3 & 0.0080 & 0.0122 & 0.0214 & \cellcolor{hcolor}{0.0077} & 0.0130 &
0.0345 & 0.0078 & 0.0150 & 0.0215 & 0.0151 & 0.0236 & 0.0236 & \textit{0.0071%
} \\
& 0.6 & 0.0560 & \cellcolor{hcolor}{0.0151} & 0.0200 & 0.0336 & 0.0221 &
0.0653 & 0.0508 & 0.0178 & 0.0325 & 0.0179 & 0.0277 & 0.0281 & \textit{0.0110%
} \\
& 0.9 & 0.4935 & 0.2730 & 0.1660 & 0.5659 & 0.7358 & 1.0847 & 0.6647 & 0.2476
& 0.2301 & 0.2553 & \cellcolor{hcolor}{0.1421} & 0.1545 & \textit{0.0034} \\
&  &  &  &  &  &  &  &  &  &  &  &  &  &
\end{tabular}%
\end{table}%

\begin{table}[tbp] \centering%
\thinmuskip=2mu\medmuskip=0mu\thickmuskip=2mu
\caption{Empirical coverage and length of nominal 95\% bootstrap HPD intervals for unadjusted and analytically-\\
bias-adjusted $\textit{LPR}(P)$ and $\textit{SPLW}(P)$, $P=0,1,2$; $T=100,\;500$, using the parametric version of the bootstrap.\\
Analogous results for intervals based on the asymptotic distribution of each of the estimators are included.\\
Figures are averaged over all values of $d$ used in the experimental design for each
value of $\phi$. \\
The highlighting indicates the empirical coverage closest to
the nominal 95\%, and the shortest length.}\label{table-hpd95}

\begin{tabular}{ccccccccccccccc}
&  &  &  &  &  &  &  &  &  &  &  &  &  &  \\[-1ex]
&  & \multicolumn{6}{c}{\underline{Panel A: $\textit{LPR}(P)$}} &  &
\multicolumn{6}{c}{\underline{Panel B: $\textit{SPLW}(P)$}} \\[2ex]
&  & \multicolumn{3}{c}{\textit{BHDCI}} & \multicolumn{3}{c}{\textit{Asymp.
interval}} &  & \multicolumn{3}{c}{\textit{BHDCI}} & \multicolumn{3}{c}{%
\textit{Asymp. interval}} \\
&  & $P=0$ & $P=1$ & $P=2$ & \multicolumn{1}{|c}{$P=0$} & $P=1$ & $P=2$ &  &
$P=0$ & $P=1$ & $P=2$ & \multicolumn{1}{|c}{$P=0$} & $P=1$ & $P=2$ \\
\cline{3-8}\cline{10-15}
&  &  &  &  &  &  &  &  &  &  &  &  &  &  \\[-0.5ex]
$\phi $ & $T$ & \multicolumn{6}{c}{Coverage} &  & \multicolumn{6}{c}{Coverage
} \\[1ex]
0.3 & 100 & 0.8828 & \cellcolor{hcolor}{0.9520} & \cellcolor{hcolor}{0.9480}
& 0.7558 & 0.8303 & 0.7845 &  & 0.8720 & 0.9603 & \cellcolor{hcolor}{0.9590}
& 0.7095 & 0.8070 & 0.7395 \\
& 500 & 0.8980 & \cellcolor{hcolor}{0.9573} & 0.9605 & 0.8348 & 0.9058 &
0.8885 &  & 0.8660 & \cellcolor{hcolor}{0.9505} & 0.9568 & 0.7878 & 0.8755 &
0.8588 \\
0.6 & 100 & 0.3210 & 0.8918 & \cellcolor{hcolor}{0.9443} & 0.1975 & 0.7113 &
0.7700 &  & 0.1865 & 0.8985 & \cellcolor{hcolor}{0.9515} & 0.0728 & 0.6700 &
0.7328 \\
& 500 & 0.2405 & 0.9220 & \cellcolor{hcolor}{0.9560} & 0.1738 & 0.8440 &
0.8735 &  & 0.0790 & 0.9310 & \cellcolor{hcolor}{0.9603} & 0.0418 & 0.8323 &
0.8655 \\
0.9 & 100 & 0.0073 & 0.2445 & \cellcolor{hcolor}{0.5923} & 0.0025 & 0.1128 &
0.3100 &  & 0.0005 & 0.1530 & \cellcolor{hcolor}{0.5298} & 0.0000 & 0.0453 &
0.2163 \\
& 500 & 0.0000 & 0.0420 & \cellcolor{hcolor}{0.4280} & 0.0000 & 0.0228 &
0.2693 &  & 0.0000 & 0.0060 & \cellcolor{hcolor}{0.2513} & 0.0000 & 0.0015 &
0.1158 \\
&  &  &  &  &  &  &  &  &  &  &  &  &  &  \\[-0.5ex]
$\phi $ & $T$ & \multicolumn{6}{c}{Interval length} &  & \multicolumn{6}{c}{
Interval length} \\[1ex]
0.3 & 100 & 0.6407 & 1.1093 & 1.5662 & \cellcolor{hcolor}{0.5016} & 0.7523 &
0.9404 &  & 0.5399 & 0.9552 & 1.3848 & \cellcolor{hcolor}{0.3911} & 0.5866 &
0.7332 \\
& 500 & 0.3274 & 0.5267 & 0.6982 & \cellcolor{hcolor}{0.2856} & 0.4283 &
0.5354 &  & 0.2629 & 0.4290 & 0.5766 & \cellcolor{hcolor}{0.2226} & 0.3340 &
0.4175 \\
0.6 & 100 & 0.6405 & 1.1039 & 1.5609 & \cellcolor{hcolor}{0.5016} & 0.7523 &
0.9404 &  & 0.5452 & 0.9570 & 1.3836 & \cellcolor{hcolor}{0.3911} & 0.5866 &
0.7332 \\
& 500 & 0.3307 & 0.5272 & 0.6983 & \cellcolor{hcolor}{0.2856} & 0.4283 &
0.5354 &  & 0.2677 & 0.4308 & 0.5778 & \cellcolor{hcolor}{0.2226} & 0.3340 &
0.4175 \\
0.9 & 100 & 0.6111 & 1.0353 & 1.4687 & \cellcolor{hcolor}{0.5016} & 0.7523 &
0.9404 &  & 0.5033 & 0.8863 & 1.2934 & \cellcolor{hcolor}{0.3911} & 0.5866 &
0.7332 \\
& 500 & 0.3306 & 0.5222 & 0.6950 & \cellcolor{hcolor}{0.2856} & 0.4283 &
0.5354 &  & 0.2648 & 0.4238 & 0.5759 & \cellcolor{hcolor}{0.2226} & 0.3340 &
0.4175 \\
&  &  &  &  &  &  &  &  &  &  &  &  &  &
\end{tabular}
\end{table}%

\begin{table}[tbp] \centering%

\thinmuskip=2mu\medmuskip=0mu\thickmuskip=2mu
\caption{Bias and mean square error (MSE) for all LPR-based estimators*; $T=100,\; 500$,
using the nonparametric version of the bootstrap, with Gaussian innovations.
Figures are averaged over the four values of $d$ used in the experimental design for each
value of $\phi$, and the lowest average bias (in absolute value) and MSE for each design highlighted.\\
Analogous figures for the (Gaussian) MLE are reported in italics.\\
(*\,Unadjusted and analytically-bias-adjusted $\textit{LPR}(P)$, $P=0,1,2$; plus bootstrap-bias-adjusted\\
$\textit{LPR}(P)\textit{-BBA}(K)$; $K=1,\ldots,3-P$, and their iteratively-adjusted (\textit{SSR}) variants.)}
\label{table-lpr-np}

\begin{tabular}{ccccccccccccccc}
&  & \multicolumn{3}{c}{$\textit{LPR}(P)$} & \multicolumn{4}{c}{$\textit{LPR}%
(0)\textit{-BBA}(K)$} & \multicolumn{3}{c}{$\textit{LPR}(1)\textit{-BBA}(K)$}
& \multicolumn{2}{r}{$\textit{LPR}(2)\textit{-BBA}(K)$} & $(Gaussian)$ \\%
[0.5ex]
&  & $P=0$ & $P=1$ & $P=2$ & \multicolumn{1}{|c}{$K=1$} & $K=2$ & $K=3$ & $%
SSR$ & \multicolumn{1}{|c}{$K=1$} & $K=2$ & $SSR$ & \multicolumn{1}{|c}{$K=1$%
} & $SSR$ & \multicolumn{1}{|c}{$ML$} \\ \hline
&  &  &  &  &  &  &  &  &  &  &  &  &  &  \\[-0.5ex]
$T$ & $\phi $ & \multicolumn{12}{c}{Bias} &  \\[1ex]
100 & 0.3 & 0.1401 & 0.0408 & 0.0161 & 0.1182 & 0.0821 & 0.0133 & 0.1034 &
0.0201 & -0.0140 & 0.0118 & \cellcolor{hcolor}{-0.0106} & -0.0260 & \textit{%
-0.1053} \\
& 0.6 & 0.3865 & 0.2022 & 0.1051 & 0.3396 & 0.2625 & 0.1237 & 0.3200 & 0.1543
& 0.0771 & 0.1418 & 0.0595 & \cellcolor{hcolor}{0.0409} & \textit{-0.0683}
\\
& 0.9 & 0.7783 & 0.7196 & \cellcolor{hcolor}{0.6485} & 0.8131 & 0.8842 &
0.9822 & 0.9018 & 0.7310 & 0.6610 & 0.7233 & 0.6577 & 0.6492 & \textit{0.0364%
} \\
500 & 0.3 & 0.0609 & 0.0118 & 0.0086 & 0.0333 & -0.0048 & -0.0656 & 0.0322 &
0.0027 & -0.0108 & 0.0034 & -0.0004 & \cellcolor{hcolor}{0.0003} & \textit{%
-0.0257} \\
& 0.6 & 0.2209 & 0.0693 & 0.0269 & 0.1545 & 0.0525 & -0.1232 & 0.1365 &
0.0270 & -0.0366 & 0.0175 & \cellcolor{hcolor}{-0.0027} & -0.0085 & \textit{%
-0.0349} \\
& 0.9 & 0.6650 & 0.4936 & 0.3709 & 0.6184 & 0.5409 & 0.3784 & 0.5833 & 0.4118
& 0.2837 & 0.3826 & 0.2866 & \cellcolor{hcolor}{0.2563} & \textit{0.0219} \\
&  &  &  &  &  &  &  &  &  &  &  &  &  &  \\[-0.5ex]
$T$ & $\phi $ & \multicolumn{12}{c}{MSE} &  \\[1ex]
100 & 0.3 & \cellcolor{hcolor}{0.0451} & 0.0747 & 0.1439 & 0.0614 & 0.1295 &
0.3644 & 0.1028 & 0.1231 & 0.2542 & 0.1584 & 0.2327 & 0.3085 & \textit{0.0552%
} \\
& 0.6 & 0.1747 & \cellcolor{hcolor}{0.1127} & 0.1511 & 0.1653 & 0.2027 &
0.4286 & 0.2045 & 0.1476 & 0.2856 & 0.1844 & 0.2409 & 0.3207 & \textit{0.0332%
} \\
& 0.9 & 0.6324 & 0.5871 & \cellcolor{hcolor}{0.5545} & 0.7363 & 1.0506 &
1.9247 & 1.2497 & 0.7027 & 0.8762 & 0.7889 & 0.7383 & 0.8570 & \textit{0.0217%
} \\
500 & 0.3 & \cellcolor{hcolor}{0.0103} & 0.0168 & 0.0300 & 0.0129 & 0.0268 &
0.0771 & 0.0145 & 0.023 & 0.0389 & 0.0282 & 0.0377 & 0.0451 & \textit{0.0096}
\\
& 0.6 & 0.0554 & \cellcolor{hcolor}{0.0214} & 0.0307 & 0.0365 & 0.0373 &
0.1258 & 0.0528 & 0.0285 & 0.0610 & 0.0500 & 0.0449 & 0.0690 & \textit{0.0134%
} \\
& 0.9 & 0.4490 & 0.2604 & 0.1677 & 0.4008 & 0.3560 & 0.3566 & 0.4387 & 0.2033
& 0.1652 & 0.2349 & \cellcolor{hcolor}{0.1384} & 0.2011 & \textit{0.0103} \\
&  &  &  &  &  &  &  &  &  &  &  &  &  &
\end{tabular}
\end{table}%

\begin{table}[tbp] \centering%

\thinmuskip=2mu\medmuskip=0mu\thickmuskip=2mu
\caption{Bias and mean square error  (MSE) for all SPLW-based estimators*; $T=100,\; 500$,
using the nonparametric version of the bootstrap, with Gaussian innovations.
Figures are averaged over the four values of $d$ used in the experimental design for each
value of $\phi$, and the lowest average bias (in absolute value) and MSE for each design highlighted.\\
Analogous figures for the (Gaussian) MLE are reported in italics.\\
(*\,Unadjusted and analytically-bias-adjusted $\textit{SPLW}(P)$, $P=0,1,2$; plus bootstrap-bias-adjusted\\
$\textit{SPLW}(P)\textit{-BBA}(K)$; $K=1,\ldots,3-P$, and their iteratively-adjusted (\textit{SSR}) variants.)}
\label{table-lpw-np}

\begin{tabular}{ccccccccccccccc}
&  & \multicolumn{3}{c}{$\textit{SPLW}(P)$} & \multicolumn{4}{c}{$\textit{%
SPLW}(0)\textit{-BBA}(K)$} & \multicolumn{3}{c}{$\textit{SPLW}(1)\textit{-BBA%
}(K)$} & \multicolumn{2}{r}{$\textit{SPLW}(2)\textit{-BBA}(K)$} & $%
(Gaussian) $ \\[0.5ex]
&  & $P=0$ & $P=1$ & $P=2$ & \multicolumn{1}{|c}{$K=1$} & $K=2$ & $K=3$ & $%
SSR$ & \multicolumn{1}{|c}{$K=1$} & $K=2$ & $SSR$ & \multicolumn{1}{|c}{$K=1$%
} & $SSR$ & \multicolumn{1}{|c}{$ML$} \\ \hline
&  &  &  &  &  &  &  &  &  &  &  &  &  &  \\[-0.5ex]
$T$ & $\phi $ & \multicolumn{12}{c}{Bias} &  \\[1ex]
100 & 0.3 & 0.1274 & \cellcolor{hcolor}{-0.0038} & -0.0375 & 0.1126 & 0.0915
& 0.0532 & 0.1124 & 0.0120 & 0.0288 & 0.0096 & -0.0194 & -0.0311 & \textit{%
-0.1053} \\
& 0.6 & 0.3907 & 0.1644 & 0.0554 & 0.3589 & 0.3081 & 0.2168 & 0.3562 & 0.1494
& 0.1216 & 0.1428 & 0.0537 & \cellcolor{hcolor}{0.0445} & \textit{-0.0683}
\\
& 0.9 & 0.7837 & 0.6950 & \cellcolor{hcolor}{0.6092} & 0.8277 & 0.8622 &
0.7982 & 0.8418 & 0.7272 & 0.6999 & 0.7122 & 0.6414 & 0.6190 & \textit{0.0364%
} \\
500 & 0.3 & 0.0572 & -0.0023 & -0.0091 & 0.0318 & -0.0023 & -0.0531 & 0.0318
& 0.0032 & 0.0081 & \cellcolor{hcolor}{0.0013} & 0.0014 & -0.0023 & \textit{%
-0.0257} \\
& 0.6 & 0.2296 & 0.0583 & 0.0105 & 0.1706 & 0.0818 & -0.0692 & 0.1627 &
0.0292 & -0.0139 & 0.0273 & 0.0016 & \cellcolor{hcolor}{0.0004} & \textit{%
-0.0349} \\
& 0.9 & 0.7150 & 0.5256 & 0.3849 & 0.7042 & 0.6922 & 0.6478 & 0.7150 & 0.4760
& 0.4006 & 0.4682 & 0.3285 & \cellcolor{hcolor}{0.3209} & \textit{0.0219} \\
&  &  &  &  &  &  &  &  &  &  &  &  &  &  \\[-0.5ex]
$T$ & $\phi $ & \multicolumn{12}{c}{MSE} &  \\[1ex]
100 & 0.3 & \cellcolor{hcolor}{0.0340} & 0.0523 & 0.1102 & 0.0379 & 0.0606 &
0.1484 & 0.0381 & 0.0770 & 0.1422 & 0.0830 & 0.1567 & 0.1966 & \textit{0.0552%
} \\
& 0.6 & 0.1720 & \cellcolor{hcolor}{0.0794} & 0.1112 & 0.1568 & 0.1540 &
0.2112 & 0.1598 & 0.0991 & 0.1677 & 0.1129 & 0.1663 & 0.1973 & \textit{0.0332%
} \\
& 0.9 & 0.6337 & 0.5343 & \cellcolor{hcolor}{0.4784} & 0.7323 & 0.9395 &
1.1430 & 0.7879 & 0.6331 & 0.7556 & 0.6613 & 0.626 & 0.6798 & \textit{0.0217}
\\
500 & 0.3 & \cellcolor{hcolor}{0.0075} & 0.0107 & 0.0197 & 0.0075 & 0.0131 &
0.0347 & 0.0075 & 0.0134 & 0.0191 & 0.0137 & 0.0225 & 0.0242 & \textit{0.0096%
} \\
& 0.6 & 0.0573 & \cellcolor{hcolor}{0.0141} & 0.0198 & 0.0355 & 0.0221 &
0.0539 & 0.0369 & 0.0175 & 0.0332 & 0.0204 & 0.0276 & 0.0281 & \textit{0.0134%
} \\
& 0.9 & 0.5169 & 0.2889 & 0.1697 & 0.5091 & 0.5227 & 0.5879 & 0.5354 & 0.2486
& 0.2084 & 0.2552 & \cellcolor{hcolor}{0.1432} & 0.1489 & \textit{0.0103} \\
&  &  &  &  &  &  &  &  &  &  &  &  &  &
\end{tabular}
\end{table}%

\begin{table}[tbp] \centering%

\thinmuskip=2mu\medmuskip=0mu\thickmuskip=2mu
\caption{Bias and mean square error (MSE) for all LPR-based estimators*; $T=100,\; 500$,
using the nonparametric version of the bootstrap, with Student \emph{t} innovations.
Figures are averaged over the four values of $d$ used in the experimental design for each
value of $\phi$, and the lowest average bias (in absolute value) and MSE for each design highlighted.\\
Analogous figures for the (Gaussian) MLE are reported in italics.\\
(*\,Unadjusted and analytically-bias-adjusted $\textit{LPR}(P)$, $P=0,1,2$; plus bootstrap-bias-adjusted\\
$\textit{LPR}(P)\textit{-BBA}(K)$; $K=1,\ldots,3-P$, and their iteratively-adjusted (\textit{SSR}) variants.)}
\label{table-lpr-st}

\begin{tabular}{ccccccccccccccc}
&  & \multicolumn{3}{c}{$\textit{LPR}(P)$} & \multicolumn{4}{c}{$\textit{LPR}%
(0)\textit{-BBA}(K)$} & \multicolumn{3}{c}{$\textit{LPR}(1)\textit{-BBA}(K)$}
& \multicolumn{2}{r}{$\textit{LPR}(2)\textit{-BBA}(K)$} & $(Gaussian)$ \\%
[0.5ex]
&  & $P=0$ & $P=1$ & $P=2$ & \multicolumn{1}{|c}{$K=1$} & $K=2$ & $K=3$ & $%
SSR$ & \multicolumn{1}{|c}{$K=1$} & $K=2$ & $SSR$ & \multicolumn{1}{|c}{$K=1$%
} & $SSR$ & \multicolumn{1}{|c}{$ML$} \\ \hline
&  &  &  &  &  &  &  &  &  &  &  &  &  &  \\[-0.5ex]
$T$ & $\phi $ & \multicolumn{12}{c}{Bias} &  \\[1ex]
100 & 0.3 & 0.1366 & 0.0501 & 0.0269 & 0.1124 & 0.0729 & -0.0026 & 0.1039 &
0.0296 & -0.0084 & 0.0198 & \cellcolor{hcolor}{-0.0016} & -0.0149 & \textit{%
-0.1054} \\
& 0.6 & 0.381 & 0.2023 & 0.1082 & 0.3304 & 0.2474 & 0.0852 & 0.3138 & 0.1501
& 0.065 & 0.1376 & 0.0584 & \cellcolor{hcolor}{0.0400} & \textit{-0.0676} \\
& 0.9 & 0.7768 & 0.7229 & 0.6499 & 0.8107 & 0.7957 & %
\cellcolor{hcolor}{0.5868} & 0.7858 & 0.7331 & 0.5965 & 0.7095 & 0.6546 &
0.6283 & \textit{0.0368} \\
500 & 0.3 & 0.0626 & 0.0169 & 0.0098 & 0.0343 & -0.0050 & -0.0675 & 0.0308 &
0.0081 & -0.0055 & 0.0041 & \cellcolor{hcolor}{0.0004} & -0.0015 & \textit{%
-0.0264} \\
& 0.6 & 0.2234 & 0.0730 & 0.0254 & 0.1560 & 0.0524 & -0.1265 & 0.1334 &
0.0305 & -0.0332 & 0.0203 & \cellcolor{hcolor}{-0.0035} & -0.0098 & \textit{%
-0.0407} \\
& 0.9 & 0.6659 & 0.4905 & 0.3687 & 0.6187 & 0.5403 & 0.3729 & 0.5747 & 0.4080
& 0.2777 & 0.3721 & 0.2840 & \cellcolor{hcolor}{0.2470} & \textit{0.0198} \\
&  &  &  &  &  &  &  &  &  &  &  &  &  &  \\[-0.5ex]
$T$ & $\phi $ & \multicolumn{12}{c}{MSE} &  \\[1ex]
100 & 0.3 & \cellcolor{hcolor}{0.0425} & 0.0753 & 0.1422 & 0.0574 & 0.1195 &
0.3435 & 0.0805 & 0.1235 & 0.2673 & 0.1681 & 0.2243 & 0.2887 & \textit{0.0554%
} \\
& 0.6 & 0.1694 & \cellcolor{hcolor}{0.1169} & 0.1519 & 0.1566 & 0.1874 &
0.3791 & 0.1893 & 0.1532 & 0.2958 & 0.1916 & 0.2417 & 0.3237 & \textit{0.0345%
} \\
& 0.9 & 0.6308 & 0.5970 & \cellcolor{hcolor}{0.5706} & 0.7358 & 0.9413 &
1.0700 & 0.8099 & 0.7198 & 0.7951 & 0.7595 & 0.7641 & 0.8416 & \textit{0.0210%
} \\
500 & 0.3 & \cellcolor{hcolor}{0.0105} & 0.0170 & 0.0292 & 0.0131 & 0.0274 &
0.0803 & 0.0183 & 0.0239 & 0.0425 & 0.0418 & 0.0368 & 0.0513 & \textit{0.0111%
} \\
& 0.6 & 0.0563 & \cellcolor{hcolor}{0.0214} & 0.0297 & 0.0370 & 0.0376 &
0.1281 & 0.0634 & 0.0282 & 0.0604 & 0.0529 & 0.0438 & 0.0692 & \textit{0.0142%
} \\
& 0.9 & 0.4503 & 0.2575 & 0.1669 & 0.4019 & 0.3586 & 0.3627 & 0.4472 & 0.2009
& 0.1647 & 0.2465 & \cellcolor{hcolor}{0.1380} & 0.2157 & \textit{0.0101} \\
&  &  &  &  &  &  &  &  &  &  &  &  &  &
\end{tabular}
\end{table}%

\begin{table}[tbp] \centering%

\thinmuskip=2mu\medmuskip=0mu\thickmuskip=2mu
\caption{Bias and mean square error  (MSE) for all SPLW-based estimators*; $T=100,\; 500$,
using the nonparametric version of the bootstrap, with Student \emph{t} innovations.
Figures are averaged over the four values of $d$ used in the experimental design for each
value of $\phi$, and the lowest average bias (in absolute value) and MSE for each design highlighted.\\
Analogous figures for the (Gaussian) MLE are reported in italics.\\
(*\,Unadjusted and analytically-bias-adjusted $\textit{SPLW}(P)$, $P=0,1,2$; plus bootstrap-bias-adjusted\\
$\textit{SPLW}(P)\textit{-BBA}(K)$; $K=1,\ldots,3-P$, and their iteratively-adjusted (\textit{SSR}) variants.)}
\label{table-lpw-st}

\begin{tabular}{ccccccccccccccc}
&  & \multicolumn{3}{c}{$\textit{SPLW}(P)$} & \multicolumn{4}{c}{$\textit{%
SPLW}(0)\textit{-BBA}(K)$} & \multicolumn{3}{c}{$\textit{SPLW}(1)\textit{-BBA%
}(K)$} & \multicolumn{2}{r}{$\textit{SPLW}(2)\textit{-BBA}(K)$} & $%
(Gaussian) $ \\[0.5ex]
&  & $P=0$ & $P=1$ & $P=2$ & \multicolumn{1}{|c}{$K=1$} & $K=2$ & $K=3$ & $%
SSR$ & \multicolumn{1}{|c}{$K=1$} & $K=2$ & $SSR$ & \multicolumn{1}{|c}{$K=1$%
} & $SSR$ & \multicolumn{1}{|c}{$ML$} \\ \hline
&  &  &  &  &  &  &  &  &  &  &  &  &  &  \\[-0.5ex]
$T$ & $\phi $ & \multicolumn{12}{c}{Bias} &  \\[1ex]
100 & 0.3 & 0.1269 & \cellcolor{hcolor}{0.0005} & -0.0218 & 0.1104 & 0.0861
& 0.0424 & 0.1093 & 0.0120 & 0.0186 & 0.0072 & -0.0054 & -0.0185 & \textit{%
-0.1054} \\
& 0.6 & 0.3914 & 0.1722 & 0.0659 & 0.3574 & 0.3040 & 0.2086 & 0.3535 & 0.1530
& 0.1196 & 0.1457 & 0.0603 & \cellcolor{hcolor}{0.0487} & \textit{-0.0676}
\\
& 0.9 & 0.7846 & 0.6982 & \cellcolor{hcolor}{0.6124} & 0.8288 & 0.8632 &
0.8055 & 0.8399 & 0.7300 & 0.6926 & 0.7124 & 0.6429 & 0.6287 & \textit{0.0368%
} \\
500 & 0.3 & 0.0576 & -0.0025 & -0.0106 & 0.0318 & -0.0026 & -0.0532 & 0.0317
& 0.0023 & 0.0066 & 0.0006 & \cellcolor{hcolor}{-0.0004} & -0.0035 & \textit{%
-0.0264} \\
& 0.6 & 0.2306 & 0.0551 & 0.0081 & 0.1706 & 0.0799 & -0.0742 & 0.1576 &
0.0254 & -0.0188 & 0.0250 & \cellcolor{hcolor}{-0.0021} & -0.0028 & \textit{%
-0.0407} \\
& 0.9 & 0.7151 & 0.5224 & 0.3799 & 0.7041 & 0.6918 & 0.6450 & 0.7142 & 0.4712
& 0.3932 & 0.4630 & 0.3210 & \cellcolor{hcolor}{0.3109} & \textit{0.0198} \\
&  &  &  &  &  &  &  &  &  &  &  &  &  &  \\[-0.5ex]
$T$ & $\phi $ & \multicolumn{12}{c}{MSE} &  \\[1ex]
100 & 0.3 & \cellcolor{hcolor}{0.0323} & 0.0525 & 0.1066 & 0.0354 & 0.0563 &
0.1386 & 0.0370 & 0.0776 & 0.1483 & 0.0931 & 0.1514 & 0.1959 & \textit{0.0554%
} \\
& 0.6 & 0.1710 & \cellcolor{hcolor}{0.0821} & 0.1134 & 0.1540 & 0.1503 &
0.2089 & 0.1590 & 0.1018 & 0.1723 & 0.1233 & 0.1656 & 0.2026 & \textit{0.0345%
} \\
& 0.9 & 0.6345 & 0.5425 & \cellcolor{hcolor}{0.4833} & 0.7325 & 0.9407 &
1.1541 & 0.7923 & 0.6480 & 0.7619 & 0.6867 & 0.6207 & 0.6530 & \textit{0.0210%
} \\
500 & 0.3 & 0.0074 & 0.0102 & 0.0200 & \cellcolor{hcolor}{0.0073} & 0.0128 &
0.0342 & 0.0074 & 0.0131 & 0.0191 & 0.0129 & 0.0225 & 0.0235 & \textit{0.0111%
} \\
& 0.6 & 0.0578 & \cellcolor{hcolor}{0.0139} & 0.0196 & 0.0360 & 0.0230 &
0.0579 & 0.0422 & 0.0174 & 0.0334 & 0.0189 & 0.0274 & 0.0276 & \textit{0.0142%
} \\
& 0.9 & 0.5168 & 0.2855 & 0.1659 & 0.5086 & 0.5209 & 0.5816 & 0.5364 & 0.2438
& 0.2023 & 0.2506 & \cellcolor{hcolor}{0.1392} & 0.1524 & \textit{0.0101} \\
&  &  &  &  &  &  &  &  &  &  &  &  &  &
\end{tabular}
\end{table}%

\end{landscape}%

\clearpage


\clearpage
\phantomsection 
\addcontentsline{toc}{section}{References}
\bibliographystyle{ims}
\bibliography{tsa}

%

\end{document}